\documentclass[12pt]{article}

\usepackage{amssymb}
\usepackage{amsmath} 
\usepackage{slashed}
\usepackage{graphicx}
\usepackage{babel}
\usepackage{hyperref}
\usepackage{color}
\usepackage[normalem]{ulem}
\usepackage{cancel}
\makeatletter
\AtBeginDocument{\let\@elt\relax}\makeatother

\usepackage{hyperref}
\usepackage{verbatim}

\newcommand{\be}{\begin{eqnarray}}
	\newcommand{\ee}{\end{eqnarray}}

\begin{document}

\begin{titlepage}

\begin{flushright}
\end{flushright}

\vspace{1ex}

\begin{center}
{\Large \bf Perturbative Analysis of CPT-Odd\\
Lorentz-Violating Scalar QCD}
\end{center}

\begin{center}
{\large J. C. C. Felipe,$^{1}$\footnote{{\tt jeanccfelipe@ufsj.edu.br}}
L. C. T. Brito,$^{2}$\footnote{{\tt lcbrito@ufla.br}}
A. C. Lehum,$^{3}$\footnote{{\tt lehum@ufpa.br}}\\
B. Altschul,$^{4}$\footnote{{\tt altschul@mailbox.sc.edu}}
A. Yu. Petrov,$^{5}$\footnote{{\tt petrov@fisica.ufpb.br}}}

\vspace{5mm}
$^{1}${\sl Departamento de Estat\' istica, F\'isica e Matem\'atica, Universidade Federal de S\~{a}o Jo\~{a}o} \\
{\sl del Rei, Rod. MG 443, Km 7, 36497-899, Ouro Branco, Minas Gerais, Brasil} \\
$^{2}${\sl Departamento de F\'{i}sica, Instituto de Ci\^{e}ncias Naturais, Universidade Federal de} \\
{\sl Lavras, Caixa Postal 3037, 37200-900, Lavras, Minas Gerais, Brasil} \\
$^{3}${\sl Faculdade de F\'{i}sica, Universidade Federal do Par\'{a}, 66075-110, Bel\'{e}m, Par\'{a}, Brazil} \\
$^{4}${\sl Department of Physics and Astronomy,} \\
{\sl University of South Carolina, Columbia, SC 29208} \\
$^{5}${\sl Departamento de F\'{i}sica, Universidade Federal da Para\'{i}ba,} \\
{\sl Caixa Postal 5008, 58051-970 Jo\~{a}o Pessoa, Para\'{i}ba, Brazil}

\end{center}

\medskip

\centerline {\bf Abstract}

\bigskip

We perform a complete one-loop renormalization analysis of CPT-odd Lorentz-violating
scalar quantum chromodynamics with adjoint scalar matter. Working to first order in the preferred background
vector and treating the corresponding operators as perturbative insertions, we compute the ultraviolet-divergent
parts of the relevant two-, three-, and four-point Green's functions for the gauge, scalar, and ghost fields.
We show that the gauge sector develops the expected Carroll-Field-Jackiw-type correction, which 
generically turns out to be
divergent in our theory, although the divergence vanishes in a certain gauge, while the scalar sector displays the
corresponding CPT-odd single-derivative term proportional to the background vector. We further demonstrate that
several of the Lorentz-violating corrections the three- and four-point function are
ultraviolet finite. All one-loop divergences may be
absorbed into counterterms already allowed by the classical Lagrangian, providing an explicit proof of the
multiplicative renormalizability of the theory at this order. We also obtain the associated renormalization
constants and one-loop $\beta$-functions for the gauge coupling, the Lorentz violation parameters, and the scalar
self-interaction.

\bigskip

\end{titlepage}

\newpage

\section{Introduction}

Symmetry has emerged as one of the most important concepts in our present understanding of
fundamental physics. The description sof the standard model and general relativity are
based on combinations of spacetime, gauge, and internal symmetry structures. Some of the symmetries
are exact, while others are only approximate, and understanding how some of the
apparent symmetries observed in low-energy effective theories may actually be merely approximate has
been very important to the development of modern physics over the last century. The standard model, which
describes what we know so far about the physics of fundamental particles, is itself normally understood to
be an effective theory, describing interactions at accessible energies in terms of
a renormalizable quantum field theory.  The symmetries that we observable in standard model physics
may themselves be only true in a low-energy approximation---meaning that those symmetries may be violated at
higher energies and more fundamental levels. So there is significant interest in the possibility of
violations of Lorentz and CPT symmetries in physics beyond the standard model. If a violation of either of
the seemingly basic symmetries is every uncovered, it will provide a important clue regarding the nature
of the most fundamental structures in physics.

An important component of the study of possible Lorentz-violating (LV) new physics is the formulation and
systematically study of generic LV models.
The standard model extension (SME)~\cite{ColKost1,ColKost2} is known to be the most natural LV framework
extending the physics we already understand.
The SME action generalizes that the standard model by introducing many additional operators
that may be constructed out of known standard model fields. These new operators resemble standard ones
structurally, but they differ that they do not need to be invariant under proper, orthochronous
Lorentz transformations. Instead, these operators may carrying free Lorentz indices, and the coefficients
multiplying the operators act as preferred vector or tensor backgrounds in spacetime.
(The SME approach has also been generalized to the include the presence of classical gravitation~\cite{KosGra}.)

The particle sector of the SME is a quantum field theory, and
full understanding of such a theory entails understanding the behavior of perturbative radiative corrections.
The SME can have quite nontrivial behavior at $\mathcal{O}(\hbar)$---such as, for example, the generation of
definitively finite yet ambiguous, regulator-dependent results for superficially divergent Lorentz- and
CPT-violating contributions to the SME's gauge field propagators.
(See Refs.~\cite{JackAmb,JK,ref-victoria2,ref-altschul37} and the further
references therein for discussions of this issue.) The standard model is a renormalizable theory, and there is
a corresponding fragment of the full SME effective field theory that is expected, for structural reasons, to be
similarly renormalizable. This minimal SME has an action that only contains operators that are
local, gauge invariant under $SU(3)_{c}\times SU(2)_{L}\times U(1)_{Y}$, and of
canonical dimension no greater than $($momentum$)^{4}$.
The most attention so far has been paid to the sectors of the SME with spinor matter, representing 
natural extensions of the spinor quantum electrodynamics (QED) that is well known as one of the most successful field
theory models, capable of making almost astonishingly precise predictions. Besides the studies of the finite
radiative corrections, the most important results for LV spinor QED
within the perturbative framework have been the explicit renormalization of the theory~\cite{KosPic}.
Various further issues related to the LV spinor QED have also been considered, both at tree level and including
loop corrections; for reviews, see Refs.~\cite{ourLV,reviewLV} and references therein.
The renormalization of LV non-Abelian gauge theories with charged gauge and spinors has also been studied,
including the case of chirally coupled fermions~\cite{ref-collad-3,Colladay:2007aj,ref-collad-1}.

At the same time, it is well known that within the SME the gauge field is coupled (as in the conventional
standard model) not just to spinor matter, but also to the scalar Higgs multiplet~\cite{ColKost1,ColKost2}.
Therefore undertaking studies of LV scalar QED and its non-Abelian generalizations is a natural task.
The first studies of LV scalar QED, with both CPT-even and CPT-odd LV operators, were presented in
Refs.~\cite{BaetaScarpelli:2003yd,Altschul:2012ig,Brito:2013npa,BaetaScarpelli:2017uxu}; the analyses included
examinations of the tree-level causality and unitarity properties of this theory, the effective potential,
and the Higgs mechanism. Further systematic calculations of quantum corrections and the renormalization of
LV scalar QED and its non-Abelian extension have also been performed~\cite{scal1,scal2,scal3,scal4}---although
these have previously been restricted to versions of the theory without CPT violation. A natural continuation of
these studies then obviously consists of considering a non-Abelian, LV gauge theory coupled to scalar matter,
along with CPT-odd terms in the action; and this is the project we have undertaken with this paper.

The structure of the paper is as follows. In section~\ref{sec-scalarQCD}, we write out the action for
our non-Abelian, LV, gauge-scalar theory---scalar quantum chromodynamics (scalar QCD)---and present its
Feynman diagram propagators and vertices.
In section~\ref{sec-gluon-self-energy}, we calculate the one-loop corrections to the quadratic action
for the gauge field, and in sections~\ref{sec-3g-vertices} and \ref{sec-4g-vertex}, respectively,
we do the same for the cubic and quartic pure gauge interaction vertices. The next three sections
deal with the analogous calculations for the scalar sector.
In section~\ref{sec-SE-scalar}, we obtain the two-point function for the scalar field.
Sections~\ref{sec-2phi-g-vertices} and \ref{sec-2phi-2g-vertices} deal with the cubic and quartic
scalar-gluon vertices, respectively, and in section~\ref{sec-4phi-vertex},
we calculate the four-point function for the scalar field. Sections~\ref{sec-SE-ghost} and \ref{sec-2c-g-vertex}
examine the ghost sector, calculating the ghost self-energy and vertex corrections, respectively.
Then section~\ref{sec-beta-functions} looks at the $\beta$-functions for the theory, and
finally, in section~\ref{sec-summary} we provide a summary and contextualization of our results.

\section{CPT-Odd LV Scalar Chromodynamics}
\label{sec-scalarQCD}

\subsection{Lagrangians for the Theory}

The scalar-gauge QFT that we shall refer to as ``scalar QCD'' is a $SU(N)$ non-Abelian gauge theory, with
the scalar matter in the adjoint representation (and thus the same representation as the gauge field itself).
Some of our results may hold for theories with gauge groups other than $SU(N)$, but examination of such
theories is beyond the scope of the present analysis.
The Lagrange density defining this theory, including possible CPT-odd LV terms is
\begin{eqnarray}
\label{LQCD-matrix}
\mathcal{L}
&=&
-\frac{1}{2}\,\mathrm{Tr}\!\left(F_{\mu\nu}F^{\mu\nu}\right)
\;+\;
\frac{1}{2\xi}\left(\partial_{\mu}G^{\mu}\right)^{2}
\;+\;
Q_2\,\kappa_\rho\,\epsilon^{\rho\sigma\mu\nu}\,\mathrm{Tr}\!\left(
G_\sigma F_{\mu\nu}
-\frac{i g}{3}\,G_\sigma [G_\mu,G_\nu]
\right)
\nonumber\\
&&
+\;2\,\mathrm{Tr}\!\left[(D_\mu\Phi)^\dagger\,D^\mu\Phi\right]
\;-\;2 m^2\,\mathrm{Tr}\!\left(\Phi^\dagger\,\Phi\right)
\;-\; i\,Q_1\,b_\mu\,\mathrm{Tr}\!\left(\Phi^\dagger\,D^\mu\Phi - (D^\mu\Phi)^\dagger\,\Phi\right)
\nonumber\\
&&
-\;\lambda_{1}\,\mathrm{Tr}\left(\Phi^\dagger\Phi\right)^{2}
\;+\;\frac{\lambda_{2}}{2}\,\mathrm{Tr}\left([\Phi^\dagger,\Phi]^{2}\right)
\;-\;\frac{\lambda_{3}}{4}\left[\mathrm{Tr}\left(\{ \Phi^\dagger,\Phi\}^{2}\right)
\;-\;\frac{1}{N_{c}}\left(\mathrm{Tr}\,\Phi^\dagger\Phi\right)^{2}\right]
\nonumber\\
&&
-\;\frac{\lambda_{4}}{4}\left[\mathrm{Tr}\left(\{\Phi^\dagger,\Phi^{\dagger}\}\,\{\Phi,\Phi\}\right)
\;-\;\frac{16}{N}\,
\mathrm{Tr}\left({\Phi^\dagger}^{2}\right)\,\mathrm{Tr}\left(\Phi^{2}\right)\right]
\;-\;2\,\mathrm{Tr}\!\left(\partial_\mu\bar c\,D^\mu c\right).
\end{eqnarray}
Here, the fields with gauge quantum numbers are $G_\mu \equiv G_\mu^a T^a$, $\Phi \equiv \phi^{a} T^a$,
and $c \equiv c^{a} T^a$; these are the gauge field itself, the scalar matter field, and the Faddeev-Popov ghost,
with $a$ being the summed gauge index in the adjoint representation.
The covariant gauge field strength is
$F_{\mu\nu} \equiv(\partial_\mu G_\nu - \partial_\nu G_\mu) - i g [\,G_\mu,\,G_\nu\,]$, while covariant derivative
of a Lie-algebra-valued field $X$ is $D_\mu X \equiv \partial_\mu X - i g [\,G_\mu,\,X\,]$. Expanding these
explicit forms, the Lagrange density may be rewritten as
\begin{eqnarray}
\label{LQCD}
\mathcal{L} & = &
- \frac{1}{4} \left(\partial^\mu {G^{\nu}}^{a} -\partial^\nu {G^{\mu}}^{a} + g f^{a b c} {G^{\mu}}^{b} {G^{\nu}}^{c} \right)
\left(\partial_\mu G_{\nu}^{a}-\partial_\nu G_{\mu}^{a} + g f^{a d e} G_{\mu}^{d} G_{\nu}^{e} \right)
+\frac{1}{2\xi}\left(\partial_{\mu}G_{a}^{\mu}\right)^{2} \nonumber \\
&& + \frac{Q_2}{2} \kappa_{\rho} \, \epsilon^{\rho \sigma \mu \nu} \left[ \frac{g}{3}
f^{a b c} G_{\mu}^{b} G_{\nu}^{c} G_{\sigma}^{a} 
+ \left(\partial^\mu G_{\nu}^{a} -\partial^\nu G_{\mu}^{a} + \frac{g}{3} f^{a d e} G_{\mu}^{d} G_{\nu}^{e}\right)
G_{\sigma}^{a} \right] \nonumber\\
&& + \left( \partial_\mu\phi^{\dagger a} + i g f^{a d e} {G^\mu}^{d} \phi^{\dagger e} \right)
\left( \partial^\mu \phi^{a} - i g f^{a b c} {G^\mu}^{b} \phi^{c} \right)
- m^2 \phi^{\dagger a}\phi^{a}  \nonumber \\
&& - \frac{i}{2} Q_1 b_\mu \left[ \phi^{\dagger a}\left( \partial^\mu \phi^{a} - i g f^{a b c} G^{\mu b} \phi^{c} \right) 
-  \left( \partial^\mu \phi^{\dagger a} + i g f^{a d e} G^{\mu d} \phi^{\dagger e} \right)\phi^{a} \right] \nonumber \\
&& 
-\frac{\lambda_1}{4}\, \big(\phi^{\dagger a} \phi^{a}\big)\,\big(\phi^{\dagger b} \phi^{b}\big)
-\frac{\lambda_{2}}{4}\, f^{abe} f^{cde}\, \big(\phi^{\dagger a}\phi^{b}\big)\,\big(\phi^{\dagger c}\phi^{d}\big)
\nonumber\\
&&
-\frac{\lambda_{3}}{4} d^{abe} d^{cde} \big(\phi^{\dagger a}\phi^{b}\big)\,\big(\phi^{\dagger c}\phi^{d}\big)
-\frac{\lambda_{4}}{4} d^{abe} d^{cde} \big(\phi^{\dagger a}\phi^{c}\big)\,\big(\phi^{\dagger b}\phi^{d}\big) \nonumber\\
&& - \bar{c}^{a} \, \partial^\mu \partial_\mu c^{a}\,+ g \partial^\mu \bar{c}^{a} \, c^{b} f^{a c b} G_{\mu}^{c}.
\end{eqnarray}

There are two sets of structure constants, $f^{abc}$ and $d^{abc}$. With the Lie algebra generators normalized so that
$\mathrm{Tr}(T^a T^b)=\tfrac{1}{2}\delta^{ab}$, they are $[T^a,T^b]=if^{abc}T^c$, the usual antisymmetric structure
constants that are also the elements of the representation matrices in the adjoint, and
$\{T^a,T^b\}=d^{abc}T^c+\frac{1}{N}\delta^{ab}{\bf 1}$, a symmetric set of structure
constants defined analogously in term of the anticommutator. The quadratic Casimir operator
in the adjoint representation is $f^{acd}f^{bcd}=C_A\delta^{ab}$, where for the gauge group $SU(N)$ we have $C_A=N$.
All contributions proportional to $C_A$ are obviously purely non-Abelian and vanish in the Abelian limit $f^{abc}=0$.

The action derived from $\mathcal{L}$ is gauge invariant under transformations with the infinitesimal forms
$\delta G_\mu = D_\mu \alpha \equiv \partial_\mu \alpha - i g [G_\mu,\alpha]$,
$\delta \Phi = -i g [\alpha,\Phi]$, and $\delta c = -i g [\alpha,c]$ for the ghosts. (The gauge parameter also
takes values in the Lie algebra, with $\alpha\equiv\alpha^a T^a$.) As is commonplace for Chern-Simons terms and
their LV generalizations, it is only the action, not $\mathcal{L}$ itself that is gauge invariant. The Lagrange
density changes by a total derivative under a gauge transformation, but this does not affect the action when
the fields vanish sufficiently rapidly at infinity. Similarly, the equations of motion involve only gauge-covariant
quantities like the field strength.

These Lagrange densities already include gauge-fixing terms, which appear in two separate ways. There is the
$\xi$-dependent term in the pure gauge sector, which is used to simplify the gauge propagator. There are also, as
noted, the Faddeev-Popov ghost terms, which arise from the exponentiation of the functional determinant of
a gauge transformation matrix in the path integrated action. There may be additional possibilities for Lorentz
violation in the ghost sector;
however, since the ghosts are, physically speaking, just another manifestation of the gauge
field---with negative norm to cancel spurious contributions from longitudinal and timelike gluons---it is expected that
such terms will generally destroy the gauge symmetry of the theory~\cite{ref-altschul25}.

There are two background vectors, $\kappa_{\mu}$ in the gauge sector---the Chern-Simons-like or Carroll-Field-Jackiw
(CFJ) term---and
$b_{\mu}$ in the scalar matter sector. These may be entirely unrelated, not pointing in the same spacetime direction.
Both of them are CPT violating, as is typical of SME backgrounds with a odd numbers of vector indices. (For quantum
field theories with well-defined $S$-matrices---meaning theories that are stable and unitary, although not
necessarily local---the strongest
expression of the CPT Theorem states that it is impossible to have physical CPT violation without
there also being Lorentz attendant violation~\cite{ref-greenberg}.) However, although they have the same behavior under
the combined CPT operation, $\kappa_{\mu}$ and $b_{\mu}$
differ in their behaviors under the individual discrete symmetry constituents. The gauge term is C even and PT odd
(with whether it is odd under P or T determined by whether a component of $\kappa_{\mu}$ is temporal or spatial,
respectively), while the scalar term is C odd and PT odd. Because of this difference, we expect that the two terms
will be forbidden from mixing under leading-order radiative corrections. While such loop corrections with $\kappa_{\mu}$
have been considered before, those involving $b_{\mu}$ are completely new.

The renormalization of this theory proceeds along
fairly standard lines. For the pure gauge sector (which necessarily includes
the ghosts fields, which are needed to preserve the unitarity of gauge interactions), including the CPT-odd
terms, the renormalization analysis has already been performed~\cite{ref-collad-3}.
However, we shall generalize the renormalization
process to include the impact of the scalar matter, which introduce interaction vertices and loop diagrams
not seen in either the pure non-Abelian gauge theory or in LV spinor QCD.

The bare fields are redefined in terms of renormalized ones,
$\phi^{a}\!\to\! Z_{2}^{1/2}\phi_{r}^{a}$, $G_{\mu}^{a}\!\to\! Z_{3}^{1/2}G_{\mu r}^{a}$ and
$c^{a}\!\to\! Z_{c}^{1/2}c_{r}^{a}$, with $Z_2$, $Z_3$, and $Z_c$ being the field strength renormalization
constants. Since we shall be treating all the Lorentz-violating terms as vertices, even when they represent
vertices with only two fields (meaning that they could be resummed as part of the exact propagators for the fields),
the field strength renormalization factors will be set entirely by the behavior of the conventions, Lorentz-invariant
kinetic terms for the various fields.
Upon substituting these redefinitions into the bare Lagrangian and, for brevity, dropping the subscript $r$
on the renormalized fields, the renormalized Lagrange density retains the form shown in eq.~\eqref{LQCD}, with the
counterterm part of $\mathcal{L}$ being given by
\begin{eqnarray}
\label{CountLag}
\mathcal{L}_{CT} &=& \delta_2\, \partial^\mu \phi_i^{\dagger a} \, \partial_\mu \phi_i^{a}
-\delta_{m^2}\, \phi^{\dagger a}\phi^{a}
-\frac{\delta_3}{2}\left[ (\partial_\nu G^{\mu a})(\partial^\nu G^a_\mu)
- (\partial_\mu G^{\nu a})(\partial_\nu G^{\mu a})\right]\nonumber\\
&&+\frac{\delta_{3}}{2\xi}\left(\partial_{\mu}G_{a}^{\mu}\right)^{2}
+ \frac{g\,\delta_{3g}}{4}\, f^{abc}\, G^b_\mu G^c_\nu ( 
\partial^\nu G^{\mu a}\, 
- \partial^\mu G^{\nu a})
 -\frac{g^2\,\delta_{4g}}{4}\,
f^{abc} f^{ade}\, {G^b}^\mu G^d_\mu {G^c}^\nu G^e_\nu
\nonumber\\
&& 
+\frac{Q_2}{2}
\kappa_\rho\, \epsilon^{\rho\sigma\mu\nu} \left[ 
\delta_{2_{CFJ}}\,
G_{\sigma}^{a} (\partial_\mu G_{\nu}^{a}\, 
- \partial_\nu G_{\mu}^{a}\,)
 + \frac{2 g}{3}~\delta_{3_{CFJ}}\,
f^{abc}\, G^b_\mu G^c_\nu G_{\sigma}^{a} \right]
\nonumber\\
&&
+ \frac{i}{2}Q_1 \delta_{Q_1}\, b_\mu\left[ 
(\partial^\mu \phi^{\dagger a}) \phi^{a}
- \phi^{\dagger a}(\partial^\mu \phi^{a})\right]
+i g\, \delta_1\, f^{abc}\, G_{\mu}^{b}\left[
\phi^{\dagger c}(\partial^\mu \phi^{a})
-i (\partial^\mu \phi^{\dagger a})\phi^{c})\right]
\nonumber\\
&&
+ g^2 \,\delta_4\,
f^{abc} f^{ade}\, G^b_\mu G^{\mu d}\, \phi^{\dagger c}\phi^{e}
 -\frac{Q_{1}g}{2}\delta_{1_{Q_1}}\,
b^\mu f^{abc}\, G^b_\mu\, \phi^{\dagger a}\phi^{c}
\nonumber\\
&& 
-\frac{\delta_{\lambda_1}}{4}\, \big(\phi^{\dagger a} \phi^{a}\big)\,\big(\phi^{\dagger b} \phi^{b}\big)
-\frac{\delta_{\lambda_{2}}}{4}\, f^{abe} f^{cde}\, \big(\phi^{\dagger a}\phi^{b}\big)\,\big(\phi^{\dagger c}\phi^{d}\big)
\nonumber\\
&&
-\frac{\delta_{\lambda_{3}}}{4} d^{abe} d^{cde} \big(\phi^{\dagger a}\phi^{b}\big)\,\big(\phi^{\dagger c}\phi^{d}\big)
-\frac{\delta_{\lambda_{4}}}{4} d^{abe} d^{cde} \big(\phi^{\dagger a}\phi^{c}\big)\,\big(\phi^{\dagger b}\phi^{d}\big)\nonumber\\
&& -\delta_c\, \bar{c}^{a} \, \partial^\mu \partial_\mu c^{a} +
g\delta_{1c}\, \partial^\mu \bar{c}^{a} \, c^{b} f^{a c b} G_{\mu}^{c}.
\end{eqnarray}
The counterterms appearing in $\mathcal{L}_{CT}$ are
\begin{eqnarray}
\label{eq4}
Z_2&=&1+\delta_2 \\
Z_3&=&1+\delta_3 \\
Z_c&=&1+\delta_c \\
Z_2m^2&=&m^2+\delta_{m^2} \\
Z_3^{3/2}g&=&g(1+\delta_{3g})  \\
Z_3^{2}g^2&=&g^2(1+\delta_{4g}) \\
Z_{2}Z_3^{1/2}\,g&=&g(1+\delta_1) \\
Z_3 Q_2 &=&Q_2(1+\delta_{2_{CFJ}}) \\
Z_3^{3/2} Q_{2}g &=& Q_{2}g(1+\delta_{3_{CFJ}}) \\
Z_2 Q_1 &=&Q_1(1+\delta_{Q_{1}}) \\
Z_2\,Z_3^{1/2}\,Q_{1}g &=& Q_{1}g(1+\delta_{1_{Q_{1}}}) \\
Z_2^2\lambda_{j}&=&\lambda_{j}+\delta_{\lambda_i} \\
Z_c\,Z_3^{1/2} g &=&g\,(1+\delta_{1c}),
\end{eqnarray}
where the index $j$ on $\lambda_{j}$ runs from $j=1\ldots 4$, covering all four scalar self-interaction terms.
From eq.~(\ref{LQCD}), we can extract the Feynman rules for the theory.

\subsection{Feynman Rules}

In this work, we are treating all the LV operators as perturbative insertions. Consequently, the
momentum-space propagators are derived
from the Lorentz-invariant quadratic part of the action in the usual manner, yielding the standard forms,
\begin{eqnarray}
\langle T~\phi^{a}(p)\phi^{\dagger b}(-p)\rangle &=& \frac{i}{p^{2}-m^{2}+i\epsilon}\delta^{ab} \\
\langle T~A^{a}_{\mu}(p)A^{b}_{\nu}(-p)\rangle &=& \frac{-i}{p^2+i\epsilon}
\left[ g_{\mu\nu}-(1-\xi)\frac{p_\mu p_\nu}{p^2}\right]\delta^{ab} \\
\langle T~c^{a}(p)\bar{c}^{b}(-p)\rangle &=& \frac{i}{p^{2}+i\epsilon}\delta^{ab}.
\end{eqnarray}
In addition, the interaction terms give rise to the following vertices:
\begin{itemize}
\item three gluon-vertex
\begin{equation}
V_{ggg}(p_{1}^{\mu},p_{2}^{\nu},p_{3}^{\alpha})=gf^{abc}\left(p_{1}^{\alpha}g^{\mu\nu}-p_{2}^{\alpha}
g^{\mu\nu}-p_{1}^{\nu}g^{\mu\alpha}+p_{3}^{\nu}g^{\mu\alpha}+p_{2}^{\mu}g^{\nu\alpha}-p_{3}^{\mu}g^{\nu\alpha}\right),
\end{equation}
\item four gluon-vertex
\begin{eqnarray}
V_{gggg}(p_{1}^{\mu},p_{2}^{\nu},p_{3}^{\alpha},p_{4}^{\beta}) & = &
ig^{2}\big(f^{ace}f^{bde}g^{\mu\beta}g^{\nu\alpha}+f^{abe}f^{cde}g^{\mu\beta}g^{\nu\alpha}
+f^{ade}f^{bce}g^{\mu\alpha}g^{\nu\beta}
\nonumber \\
& & -f^{abe}f^{cde}g^{\mu\alpha}g^{\nu\beta}
-f^{ade}f^{bce}g^{\mu\nu}g^{\alpha\beta}-f^{ace}f^{bde}g^{\mu\nu}g^{\alpha\beta}\big)\!,
\end{eqnarray}
\item two-gluon LV-vertex 
\begin{equation}
V_{gg\kappa}(p_{2}^{\mu},p_{3}^{\nu})=\frac{Q_{2}\kappa_{\alpha}\delta_{bc}}{2}\left[\epsilon^{\mu\nu\alpha}{}_{\beta}
(p_{2}-p_{3})^{\beta}+\epsilon^{\mu\nu\alpha}{}_{\lambda}(p_{2}-p_{3})^{\lambda}\right],
\end{equation}
\item three-gluon LV vertex
\begin{equation}
V_{ggg\kappa}=-2igQ_{2}\kappa_{\beta}\epsilon^{\mu\nu\alpha\beta}f^{bcd},
\end{equation}
\item one-gluon, two-ghost vertex
\begin{equation}
V_{g \bar{c}c}(p_{1}^{\mu})=-gf^{cab}p_{1}^{\mu},
\end{equation}
\item two-scalar LV-vertex
\begin{equation}
V_{\phi^{\dagger}\phi b}(p_{2},p_{3})=\frac{iQ_{1}b_{\mu}}{2}(p_{3}-p_{2})^{\mu},
\end{equation}
\item one-gluon, two-scalar vertex
\begin{equation}
V_{g\phi^{\dagger}\phi}(p_{2},p_{3})=gf^{cab}\left(p_{3}-p_{2}\right)^{\mu},
\end{equation}
\item one-gluon, two-scalar LV-vertex
\begin{equation}
V_{g\phi^{\dagger}\phi b}=gQ_{1}f^{bcd}b^{\mu},
\end{equation}
\item two-gluon, two-scalar-vertex
\begin{equation}
V_{gg\phi^{\dagger}\phi b}=ig^{2}g^{\mu\nu}\left(f^{ade}f^{bce}+f^{ace}f^{bde}\right),
\end{equation}
\item four-scalar vertex
\begin{eqnarray}
V_{\phi^{\dagger}\phi^{\dagger}\phi\phi} &=& \frac{-i}{2} \big(\lambda_{1}\delta^{ad}\delta^{bc}
+\lambda_{1}\delta^{ac}\delta^{bd}+\lambda_{2}f^{ade}f^{bce}+\lambda_{2}f^{ace}f^{bde} \nonumber\\
&& +\lambda_{3}d^{ade}d^{bce}+\lambda_{3}d^{ace}d^{bde}+\lambda_{4}\delta^{ad}\delta^{bc}
+\lambda_{4}\delta^{ac}\delta^{bd}\big).
\end{eqnarray}
\end{itemize}
Having derived the Feynman rules, we can proceed to calculate the radiative contributions to the
two-, three- and four-point functions at the one-loop order.

\section{Calculations of the One-Loop Gluon Self-Energy}
\label{sec-gluon-self-energy}

Using the Feynman rules given above, we can calculate the one-loop contributions to the correlations functions
of the theory. As well as working to $\mathcal{O}(\hbar)$, we shall also work only to lowest order in the
CPT-violating backgrounds $b_{\mu}$ and $\kappa_{mu}$---by which we meant neglecting terms at
$\mathcal{O}(b^{2})$, $\mathcal{O}(\kappa^{2})$, $\mathcal{O}(b\kappa)$, and higher.
We shall  start with the field strength renormalization for the gauge field.

\subsection{Zeroth Order in the LV Vectors}
\label{sec-SE-gluon-zeroth}

\begin{figure}[h!]
    \centering
   \includegraphics[width=12cm, height=4cm]{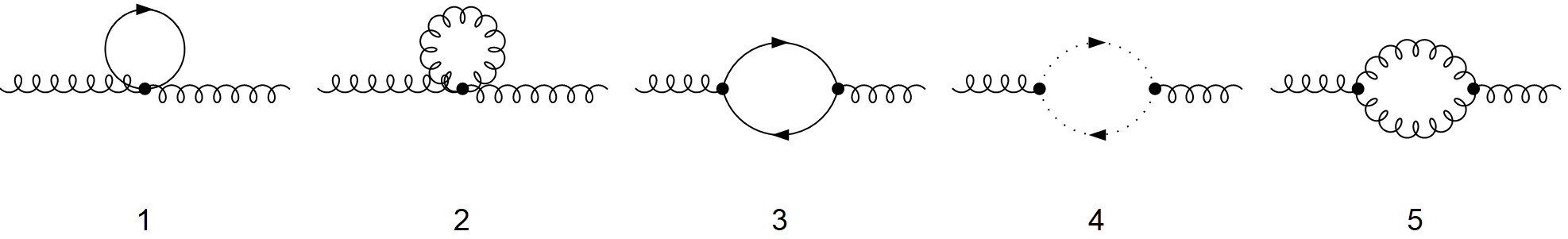}
    \caption{One-loop contributions to the gluon self-energy. Solid, curly, and dotted lines denote scalar, gluon,
    and ghost propagators, respectively.}
    \label{Fig1}
\end{figure}

Our first step in the calculation of the wave function renormalization for the gluon field consists
of verifying the Lorentz-invariant part of the two-point function for the gauge field. Here, as well as below, we consider
all three types of loops---gauge, scalar and, ghost ones.
The five corresponding Feynman diagrams are show in fig.~\ref{Fig1}. For each type of field there is a vacuum
polarization diagram that depends on the momentum of the propagating gauge field, and for the gauge and scalar fields
there are also momentum-independent tadpole diagrams that are necessary to maintain the gauge invariance
of the renormalized action (by canceling the quadrating divergences).

To perform this and subsequent calculations, we used a suite of Mathematica
pack\-ages~\cite{Mertig1991,Hahn1999,Hahn2001,Mertig2016,Mertig2023}. These computational tools facilitate the
computation and manipulation of Feynman diagrams and associated quantities. In our study, we
concentrate on the portions of the various diagrams
that are divergent in the ultraviolet (UV), since these are what is necessary for understanding
the renormalization of the theory (although this is not to say there are not extremely interesting phenomena also
to be found among the UV-finite radiative corrections). Explicitly, we compute the 
sum of the infinite parts of the five diagrams depicted in fig.~\ref{Fig1}, which only contribute to the wave
function renormalization of the gauge field. Divergences are handled using dimensional regularization, with
the presence of the two tadpole diagrams ensuring that there is no quadratic divergence, corresponding to a factor of
$\Gamma(1-d/2)$. The resulting gluon self-energy tensor can be written as
\begin{equation}
\Pi_{\mu\nu}^{ab}(p)=
-\frac{i\, g^{2}C_{A}\left(11-3\xi\right)}{96\,\pi^2\,\epsilon}
\left(g_{\mu\nu}\, p^2- p_\mu p_\nu \right)
\delta^{ab} \, +\mathrm{finite}.
\end{equation}
This polarization tensor is transverse to the gauge field energy-momentum vector, as it must be.
The quantity $(11-3\xi)$ represents $\left(13-3\xi-2N_{s}\right)$,
where $N_{s}$ is the number of scalar fields in the adjoint representation and we have taken $N_{s}=1$.
The finite part of the self-energy may be written in terms of scalar Passarino-Veltman integrals.

\subsection{First Order in the Lorentz Violation}
\label{sec-SE-gluon-first}

\begin{figure}[h!]
    \centering
   \includegraphics[width=12cm, height=4cm]{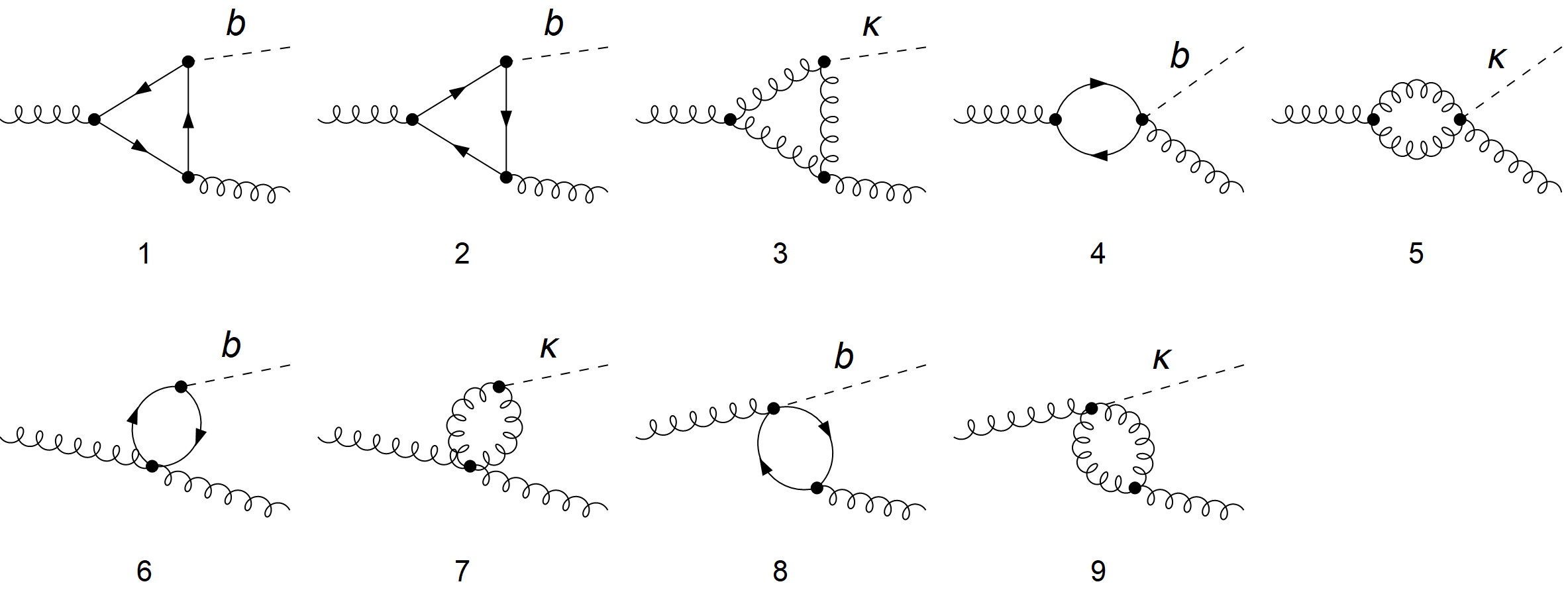}
    \caption{One-loop contributions to the gluon self-energy at first order in the LV insertion.
    The dashed line denotes a LV insertion, through which no energy-momentum is exchanged.}
    \label{Fig2}
\end{figure}

The Feynman diagrams contributing to the gluon self-energy at first order in the LV insertions
$\kappa_{\mu}$ and $b^{\mu}$ are shown in fig.~\ref{Fig2}.

Summing these diagrams, the LV correction takes the form
\begin{equation}
\Pi_{\mu\nu}^{ab}(p)
=\frac{Q_2 \, g^{2}C_{A}(3+\xi)}{16\,\pi^2\,\epsilon}\,
\epsilon_{\mu\nu\alpha\beta}\, \kappa^\alpha p^\beta \,
\delta^{ab} + \text{finite}. 
\end{equation}
This result corresponds in form precisely to bilinear part of the CFJ term (as expressed in momentum space).
Therefore, we have confirmed the presence of a divergent CFJ self-renormalization. It is interesting that the
contribution from non-Abelian self-interactions is indeed divergent, while contributions from spinor matter are --- while
they involve superficially divergent integrals --- always manifestly finite, because of cancellations between
right- and left-chiral virtual intermediaries. Moreover, as expected there are no contribu- tions from the diagrams
with $b_{\mu}$ insertions, because they have the wrong behavior under C. Moreover, the loop correction is
obviously essentially non-Abelian, vanishing in the Abelian case $C_A=0$.

\subsection{Counterterms for the Gluon Self-Energy}

Taking the results from sections~\ref{sec-SE-gluon-zeroth} and \ref{sec-SE-gluon-first}, we can write down the
counterterms necessary in the quadratic part of the gauge Lagrange density.

\begin{figure}[h!]
    \centering
   \includegraphics[width=4cm, height=2cm]{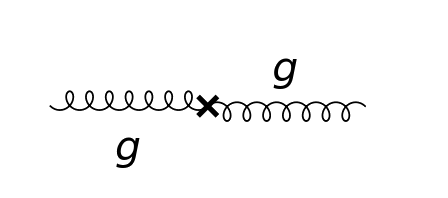}
    \caption{Counterterm for the Lorentz-invariant gluon two-point function.}
    \label{FigCT1}
\end{figure}
The counterterms associated with the Lorentz-invariant gluonic sector are represented diagrammatically in fig.~\ref{FigCT1} and are given by
\begin{equation}\label{eq:delta3}
\delta_3 = 
\frac{g^{2}C_A \left(11-3\xi\right)}
{96\,\pi^2\,\epsilon}.
\end{equation}
%
%
\begin{figure}[h!]
    \centering
   \includegraphics[width=4cm, height=2cm]{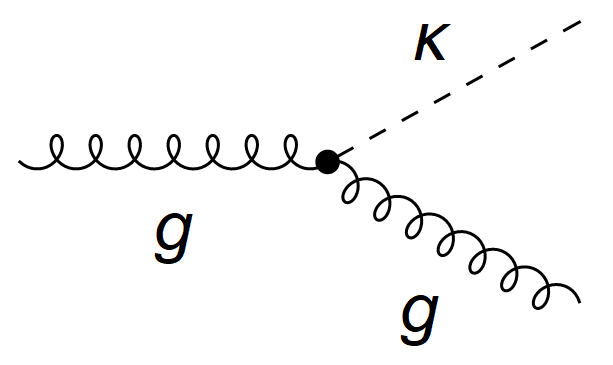}
    \caption{Counterterm for the gluon two-point vertex at first order in the LV insertion.}
    \label{FigCTLV1}
\end{figure}

The expression for the counterterm corresponding to fig.~\ref{FigCTLV1}, which is intrinsically first order
in the Lorentz violation parameter $\kappa_{\mu}$, is
\begin{equation}\label{2CFJ}
\delta_{2_{CFJ}} = 
-\frac{g^{2}C_{A}(3+\xi)}{32\,\pi^2\,\epsilon}.
\end{equation}
These renormalization constants will be utilized further
when we examine the re\-nor\-mal\-i\-za\-tion group equations. We note that this result does not contradict to the
Abelian result obtained in Ref.~\cite{scal4}, in which the one-loop contribution to the CFJ term was shown to vanish,
since in the present case, the coupling between the gauge field and the matter involves their commutators, instead of
the simple products that are present in the Abelian Lagrange density. This is tied to the fact that in the Abelian
theory $C_A=0$; hence the result \eqref{2CFJ} reduces to zero in that case.
We note, however, that the divergent part of the CFJ term vanishes even in a non-Abelian case at a certain value
of the gauge parameter, namely, $\xi=-3$~\cite{ref-collad-3}. (Note that this is not the same choice as the Yennie
gauge---useful for eliminating some infrared divergences in gauge field self-energy calculations---which has $\xi=+3$.)

\section{Calculations of Three-Gluon Vertices}
\label{sec-3g-vertices}

Now we turn to the evaluation of
quantum corrections to the interaction vertices in the gluon sector, beginning with the trilinear vertex.

\subsection{Three-Gluon Vertex at Zeroth Order in the LV Parameters}


\begin{figure}[h!]
    \centering
   \includegraphics[width=14cm, height=6cm]{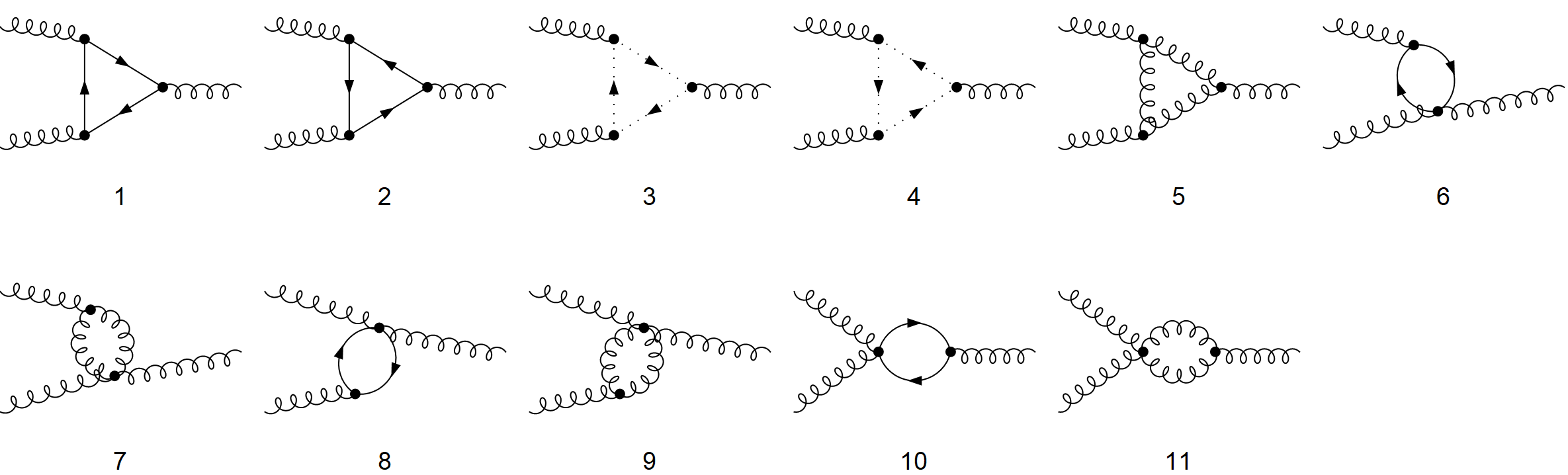}
    \caption{One-loop contributions to the three-gluon vertex.}
    \label{Fig4}
\end{figure}

The one-loop corrections to this vertex are shown in fig.~\ref{Fig4}. Isolating the UV-divergent parts
of these diagrams and summing all contributions, we find
\begin{eqnarray}
\Gamma_{\mu\nu\alpha}^{abc}(p_1,p_2,p_3) &=& -\frac{g^{3}C_A}{192\pi^{2}\epsilon}f^{abc}
(9\,p_{1\alpha} g_{\mu\nu}
-17\,p_{2\alpha} g_{\mu\nu}
+18 \xi\, p_{2\alpha} g_{\mu\nu}
+4\,p_{3\alpha} g_{\mu\nu}
-9 \xi\, p_{3\alpha} g_{\mu\nu}
\nonumber\\
&&
+\,9\,p_{1\nu} g_{\alpha\mu}
+4\,p_{2\nu} g_{\alpha\mu} 
-9 \xi\,p_{2\nu} g_{\alpha\mu}
-17\,p_{3\nu} g_{\alpha\mu}
+18 \xi\,p_{3\nu} g_{\alpha\mu} \nonumber\\
&&
+\,13\,p_{2\mu} g_{\alpha\nu}
-9 \xi\,p_{2\mu} g_{\alpha\nu}
+13\,p_{3\mu} g_{\alpha\nu}
-9 \xi\,p_{3\mu} g_{\alpha\nu}),
\end{eqnarray}
which can be shown (using energy-momentum conservation at the vertex)
to reproduce the Lorentz and gauge index structures of the three-point vertex in the gauge sector.
We note again that this contribution vanishes in the Abelian case where $f^{abc}=0$ and $C_A=0$.

\subsection{Three-Gluon Vertex at First Order}

\begin{figure}[h!]
    \centering
   \includegraphics[width=16cm, height=8cm]{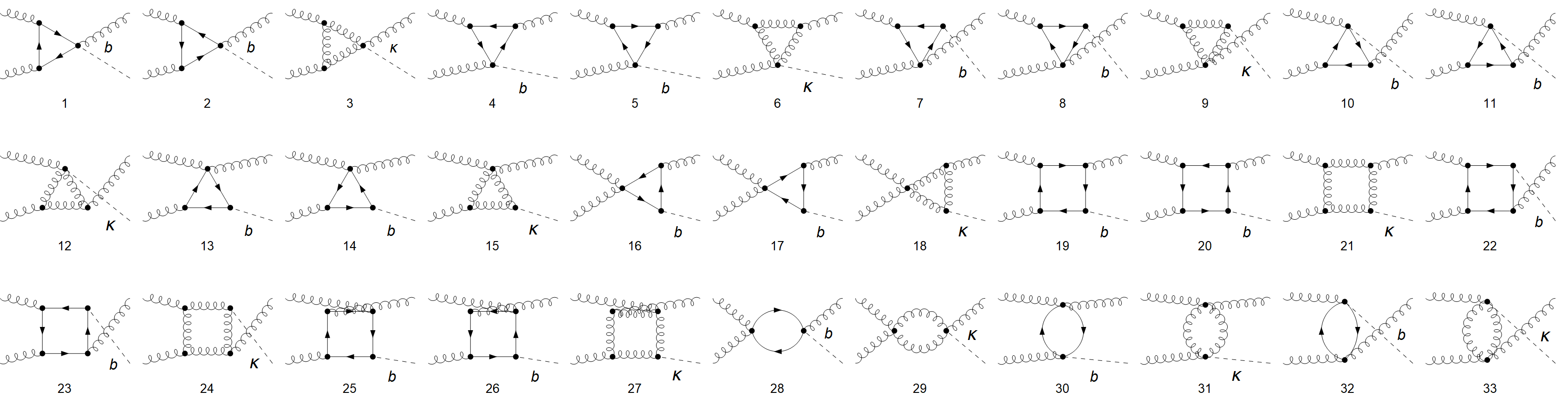}
    \caption{One-loop contributions to the three-gluon vertex at first order in the LV insertion.}
    \label{Fig5}
\end{figure}

The Feynman diagrams of third order in the gauge field and the first order in the LV vectors
are enumerated in fig.~\ref{Fig5}.

After a lengthy calculation, the sum of all diagrams is found to be
\begin{equation}
\label{eqCFJ23}
\Gamma_{\mu\nu\alpha}^{abc}(p_1,p_2,p_3) \;=\;
-\frac{3i \, Q_{2} g^{3} C_{A}(3+\xi)}{32\,\pi^2\,\epsilon} \,
\epsilon_{\mu\nu\alpha\beta}\, \kappa^{\beta} \,
f^{abc},
\end{equation}
which is a radiative correction to the self-interaction part of the CFJ term.
It is clear that this result vanishes for an Abelian group where $f^{abc}=0$ and $C_A=0$, as well as for
an arbitrary gauge group with the gauge choice $\xi=-3$.

A comment is in order here. The constant $Q_{2}$, shown in eq.~\eqref{eqCFJ23}, controls the overall
normalization of the CFJ term, as we can see from the counterterm Lagrange density \eqref{CountLag},
while the term $\delta_{2_{CFJ}}$ renormalizes its two-gluon part term and $\delta_{3_{CFJ}}$ its three-gluon
self-interaction term. These terms are specified in the eqs.~\eqref{2CFJ} and \eqref{3CFJ}, respectively. They
have the correct
relative sizes to preserve the gauge-invariant Chern-Simons-like form for the CFJ term in the integrated action.

\subsection{Counterterms to the cubic vertex}

Using the results of the previous subsections, we
may write down our counterterms for the cubic vertex, involving term of both the
zeroth and first orders in $\kappa_{\mu}$.

\begin{figure}[h!]
    \centering
   \includegraphics[width=4cm, height=2cm]{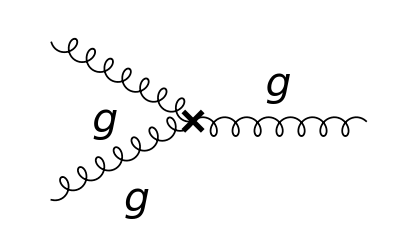}
    \caption{Counterterm for the Lorentz-invariant three-gluon vertex.}
    \label{CTFig2}
\end{figure}

The expression for the counterterm of zeroth order in the LV parameter, depicted in fig.~\ref{CTFig2}, is
\begin{equation}\label{eq:delta3g}
\delta_{3g} = 
\frac{g^{2}C_{A}\,(13-9\xi)}
{192\,\pi^2\,\epsilon},
\end{equation}
whereas with multiple scalar fields, $(13-9\xi)$ would be replaced by $(17-9\xi-4N_{s})$.

\begin{figure}[h!]
    \centering
   \includegraphics[width=4cm, height=2cm]{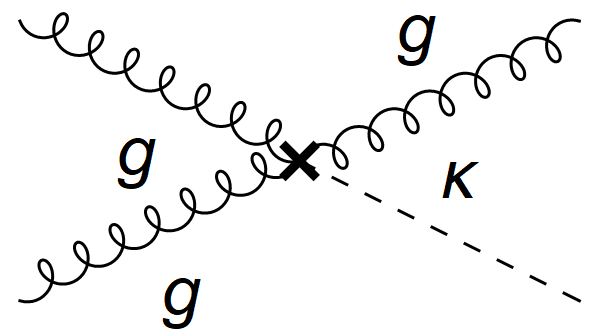}
    \caption{Counterterm for the three-gluon vertex at first order in the LV insertion.}
    \label{CTLVFig2}
\end{figure}

The expression for the counterterm of the first order in the LV parameter depicted in fig.~\ref{CTLVFig2} is
\begin{equation}\label{3CFJ}
\delta_{3_{CFJ}}= 
-\frac{3g^{2}C_{A} \, (3+\xi)}
{64\, \pi^2\, \epsilon}.
\end{equation}
As already noted, this contribution vanishes at $\xi=-3$, just as does the corresponding counterterm 
for the the two-point function. Both quadratic and cubic parts of our CFJ
term are renormalized with the same multiplicative renormalization constant---as
they must be for reasons of gauge symmetry. The multiplicative constant depends on both the
gauge group (vanishing in the Abelian case) and the choice of gauge.

\section{Calculation of the Four-Gluon Vertex}
\label{sec-4g-vertex}

\subsection{Lorentz-Invariant Four-Gluon Divergence}

\begin{figure}[h!]
    \centering
   \includegraphics[width=16cm, height=10cm]{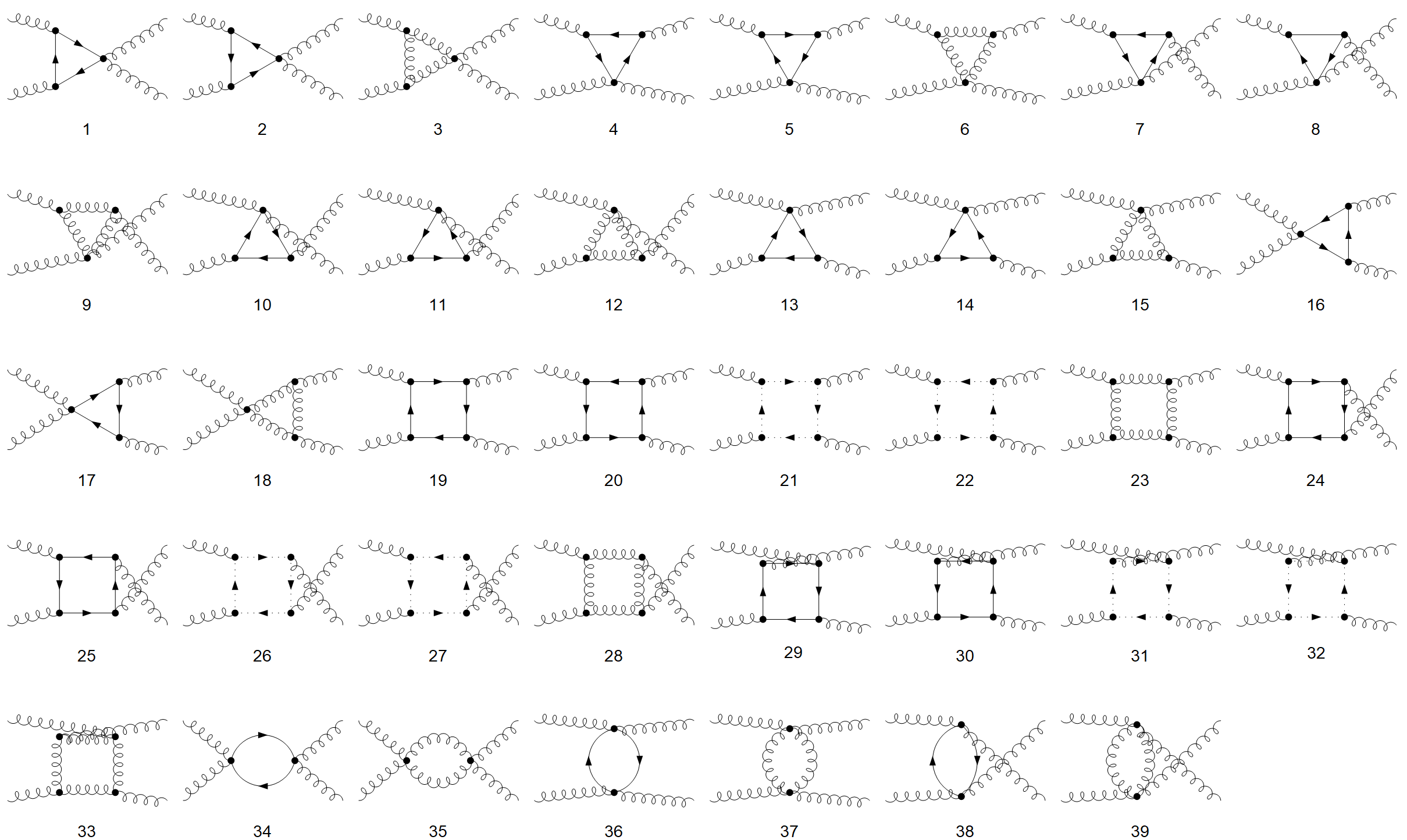}
    \caption{One-loop contributions to the Lorentz-invariant four-gluon vertex.}
    \label{Fig6}
\end{figure}

Having addressed the renormalization of the CFJ term, which involved term of both second and third orders in
the gluon fields, we shall now move on to the vertices that arise at fourth order in these fields.
We start, as previously, at zeroth order in the LV parameter.
All the diagrams contributing to the
standard one-loop four-gluon vertex in scalar QCD are shown in fig.~\ref{Fig6}.

After summing all diagrams and retaining only the divergent parts (which means that no quantities except the
loop momentum need to retained in the denominators of the Feynman propagators, since the whole
four-boson expression is only superficially logarithmically divergent), we obtain
\begin{eqnarray}
\mathcal{M}_{\alpha\mu\beta\nu}^{abcd}
&=&
\frac{i\, g^{4}C_{A}}{24\,\pi^2\,\epsilon}\,(1-3\xi)\,
\left[
-2\, g_{\alpha\mu} g_{\beta\nu}\, \mathrm{Tr}(T^a T^b T^c T^d)
+    g_{\alpha\nu} g_{\beta\mu}\, \mathrm{Tr}(T^a T^b T^c T^d)\right.
\nonumber\\
&&
+\,g_{\alpha\beta} g_{\nu\mu}\, \mathrm{Tr}(T^a T^b T^c T^d)
+    g_{\alpha\mu} g_{\beta\nu}\, \mathrm{Tr}(T^a T^b T^d T^c)
+    g_{\alpha\nu} g_{\beta\mu}\, \mathrm{Tr}(T^a T^b T^d T^c)
\nonumber\\
&&
-\, 2\, g_{\alpha\beta} g_{\nu\mu}\, \mathrm{Tr}(T^a T^b T^d T^c)
+    g_{\alpha\mu} g_{\beta\nu}\, \mathrm{Tr}(T^a T^c T^b T^d)
- 2\, g_{\alpha\nu} g_{\beta\mu}\, \mathrm{Tr}(T^a T^c T^b T^d)
\nonumber\\
&&
+\,    g_{\alpha\beta} g_{\nu\mu}\, \mathrm{Tr}(T^a T^c T^b T^d)
+    g_{\alpha\mu} g_{\beta\nu}\, \mathrm{Tr}(T^a T^c T^d T^b)
+    g_{\alpha\nu} g_{\beta\mu}\, \mathrm{Tr}(T^a T^c T^d T^b)
\nonumber\\
&&
-\, 2\, g_{\alpha\beta} g_{\nu\mu}\, \mathrm{Tr}(T^a T^c T^d T^b)
+    g_{\alpha\mu} g_{\beta\nu}\, \mathrm{Tr}(T^a T^d T^b T^c)
- 2\, g_{\alpha\nu} g_{\beta\mu}\, \mathrm{Tr}(T^a T^d T^b T^c)
\nonumber\\
&&
+\,    g_{\alpha\beta} g_{\nu\mu}\, \mathrm{Tr}(T^a T^d T^b T^c)
- 2\, g_{\alpha\mu} g_{\beta\nu}\, \mathrm{Tr}(T^a T^d T^c T^b)
+    g_{\alpha\nu} g_{\beta\mu}\, \mathrm{Tr}(T^a T^d T^c T^b)
\nonumber\\
&&
+ \left. g_{\alpha\beta} g_{\nu\mu}\, \mathrm{Tr}(T^a T^d T^c T^b)\right].
\end{eqnarray}
The UV pole of the four–gluon amplitude may be identified by projecting onto the same Lorentz-color tensor structure
as the tree-level vertex---namely the combinations of $g_{\alpha\mu} g_{\beta\nu}$, $g_{\alpha\beta} g_{\mu\nu}$,
and $g_{\alpha\nu} g_{\beta\mu}$, multiplied by the corresponding traces of the color generators,
together with all their Bose-symmetric permutations.

\subsection{LV Four-Gluon Term}

In the present calculation, we have retained only the UV-divergent parts of the diagrams. Looking at the integrals
themselves, it may appear remarkable that all the contributions --- 171 diagrams in total --- together vanish identically.
This outcome is in complete agreement with the multiplicative renormalizability of the theory and can, in fact,
be anticipated from simple considerations of power-counting and gauge symmetry~\cite{ref-collad-3}.

\subsection{Counterterm for the Four-Gluon Vertex}

\begin{figure}[h!]
    \centering
   \includegraphics[width=4cm, height=2cm]{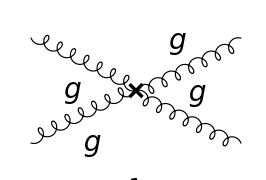}
    \caption{Counterterm for the Lorentz-invariant four-gluon vertex.}
    \label{CTFig7}
\end{figure}

Because there is no divergent contribution to the gluon four-point amplitude,
the expression for the counterterm vertex depicted in the fig.~\ref{CTFig7} is the standard one,
\begin{equation}\label{eq:delta4g}
\delta_{4g} \;=\;
\frac{g^{2}C_{A}\, (1-3\xi)}{48\, \pi^2\, \epsilon},
\end{equation}
obtained by matching the coefficient of the $1/\epsilon$ pole to the tree-level structure.

\section{Calculations of the One-Loop Scalar Self-Energy}
\label{sec-SE-scalar}

\subsection{Lorentz-Invariant Scalar Field Self-Energy}

\begin{figure}[h!]
    \centering
   \includegraphics[width=12cm, height=4cm]{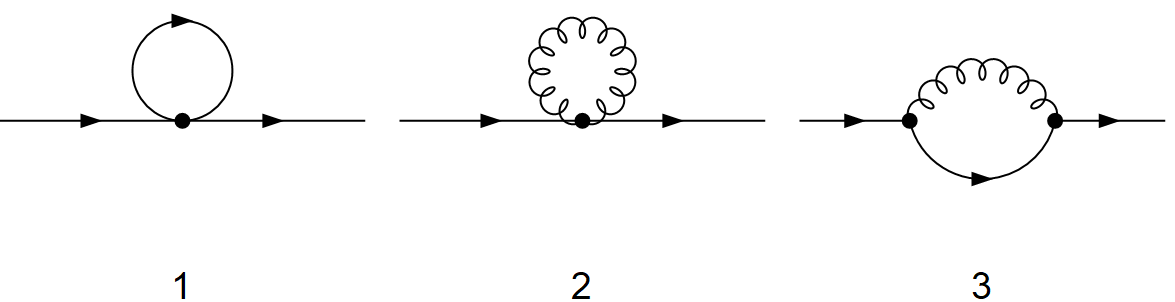}
    \caption{One-loop conventional contributions to the scalar self-energy.}
    \label{Fig11}
\end{figure}

Having dealt with the complete renormalization of the pure gauge sector of our theory,
we shall now turn to the scalar sector.
The scalar self-energy diagrams of zeroth order in $b_{\mu}$ are shown in fig.~\ref{Fig11}.

Again, these diagrams have a naive quadratic degree of divergence, and unlike for the gluon self-energy, there is
not a complete cancellation of the divergences at this order.
One of the quadratically divergent terms involves scalar self-couplings $\lambda_{j}$,
while the other two involve the gauge coupling $g$. These are independent parameters, not connected by any symmetry,
so there are not expected to be cancellations between their
contributions without fine tuning. Moreover, even with such fine tuning at
one scale, the coupling constants involved run at different rates under the renormalization group, so the finely
tuned cancellation will not persist at other scales. It is known that quantum corrections from a gauge
interaction may change whether there is spontaneous symmetry breaking in the scalar sector.  (See, for example,
the Coleman-Weinberg model~\cite{ref-coleman-weinberg} for the Abelian version.)
In general, it is not possible for coupled scalar fields to be free from radiative corrections to their masses,
unless there is a protective symmetry involved. This could mean either a spontaneously broken symmetry, which
protects the Goldstone bosons from acquiring masses, or a supersymmetry connected the scalars to fermions whose
masslessness is protected by chiral symmetry.

However, we shall not discuss the mass renormalization in this theory further, since it turns out that at
one loop order, the LV interactions play now role at lowest order. The mass counterterm is not affected and remains
the standard one for scalar QCD. Moreover, if we are primarily interested in high-energy phenomena, well above the
scalar mass scale, the only role of the mass would be as an infrared regulator, which might be removed at the end of
a calculation. With these observations in mind, the expression for the sum of diagrams in fig.~\ref{Fig11} is
\begin{equation}
\Pi_{S}^{ab}(p) \;=i\delta^{ab}\left[\delta_{m^{2}}\,
-\,\frac{g^{2}C_{A}}{16\,\pi^2\,\epsilon}\,
(3-\xi)\, p^2 \right] +\mathrm{finite}.
\end{equation}
Besides the mass term, which is infinite with a $1/(\epsilon-2)$ singularity but independent of $p$, this is 
just the standard renormalization of the $\phi^{\dagger}\Box\phi$ term in the scalar QCD action.
It is interesting to note that the field strength renormalization part
vanishes for a certain value of the gauge parameter, namely, $\xi=3$
(corresponding, as already noted, to the Yennie gauge from QED).

\subsection{First Order in the LV Parameter}

\begin{figure}[h!]
    \centering
   \includegraphics[width=14cm, height=4cm]{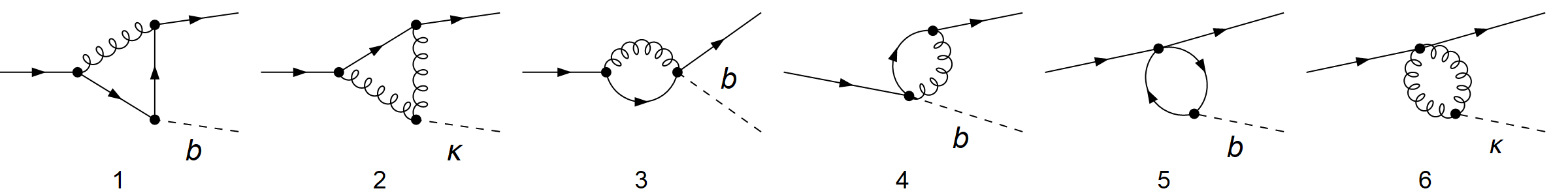}
    \caption{One-loop contributions to the scalar self-energy at first order in the LV insertion.}
    \label{Fig12}
\end{figure}

The scalar self-energy diagrams with a single LV insertion are shown in fig.~\ref{Fig12}. 
Their total contribution reads
\begin{eqnarray}
\Pi^{ab}_{S}(p) &=&
\frac{i\, Q_{1}g^{2}C_{A}}{16\,\pi^2\,\epsilon}\,
(3-\xi)\,\delta^{ab}\,(b\cdot p)\;+\;\text{finite}.
\end{eqnarray}
This result confirms that the LV insertion generates a divergent contribution to the scalar self-energy
that is proportional to $b \cdot  p$, and which thus can be consistently absorbed by the corresponding counterterm.
As expected, based on C-parity differences, there is no mixing from the diagrams with $\kappa_{\mu}$ insertions.

\subsection{Scalar Self-Energy Counterterms}

\begin{figure}[h!]
    \centering
   \includegraphics[width=6cm, height=2cm]{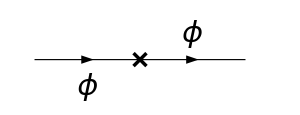}
    \caption{Counterterm for the Lorentz-invariant scalar two-point function.}
    \label{Fig32}
\end{figure}

From the divergent parts of the scalar self-energy diagrams, we extract the following counterterms in the \emph{MS} scheme.  
The wave-function renormalization counterterm for the scalar field, depicted in fig.~\ref{Fig32}, is given by
\begin{equation}
\label{eq:delta2}
\delta_{2} \;=\; 
\frac{g^{2}C_{A}\,(3-\xi)}{16\,\pi^2\,\epsilon}.
\end{equation}

\begin{figure}[h!]
    \centering
   \includegraphics[width=4cm, height=2cm]{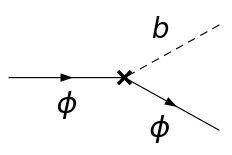}
    \caption{Counterterm for the scalar two-point function at first order in the LV insertion.}
    \label{Fig33}
\end{figure}

The mass counterterm associated with the LV contribution is zero. Finally, the renormalization of the LV coupling $Q_1$ is controlled by
\begin{equation}
\label{eq:deltaQ1}
\delta_{Q_1} \;=\;
\frac{g^{2}C_{A}\, (3-\xi)}{16\,\pi^2\,\epsilon},
\end{equation}
which is shown in fig.~\ref{Fig33}.

\section{Calculations of the Three-Field Scalar-Gluon Vertex}
\label{sec-2phi-g-vertices}

Now, let us study quantum corrections to scalar-gluon vertices.

\subsection{Lorentz-Invariant Scalar-Scalar-Gluon Vertex}

\begin{figure}[h!]
    \centering
   \includegraphics[width=12cm, height=2cm]{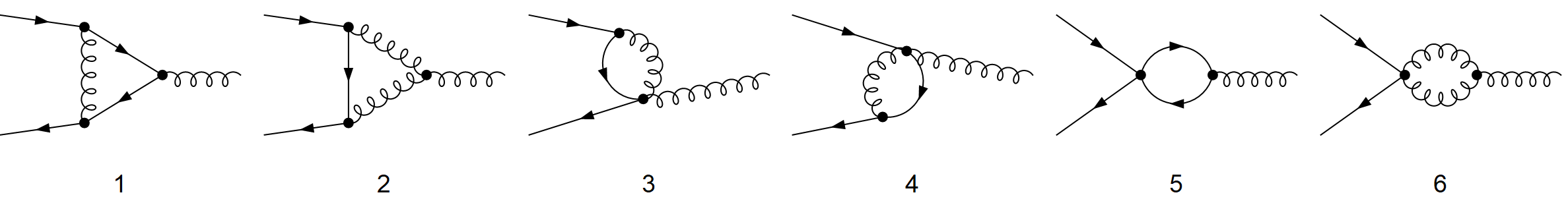}
    \caption{One-loop contributions to the scalar-scalar-gluon vertex.}
    \label{Fig14}
\end{figure}

We start with the three-point functions. The corresponding diagrams, to zeroth order in the LV parameters $b_{\mu}$
and $\kappa_{\mu}$, are shown in fig.~\ref{Fig14}.

Taking into account only the divergent parts, the result of the sum of al these diagrams turns out to be given by
\begin{equation}\label{ssg1}
\Gamma_{\mu}^{a b c}(p_1,p_2)=
\frac{g^{3}C_{A}\,(9-5\xi)}{64\,\pi^{2}\,\epsilon}
\left(p_{1\mu} - p_{2\mu}\right)\,
f^{a b c},
\end{equation}
which is consistent with the gauge invariance requirement.

\subsection{Scalar-Scalar-Gluon Vertex at First Order}

\begin{figure}[h!]
    \centering
   \includegraphics[width=16cm, height=4cm]{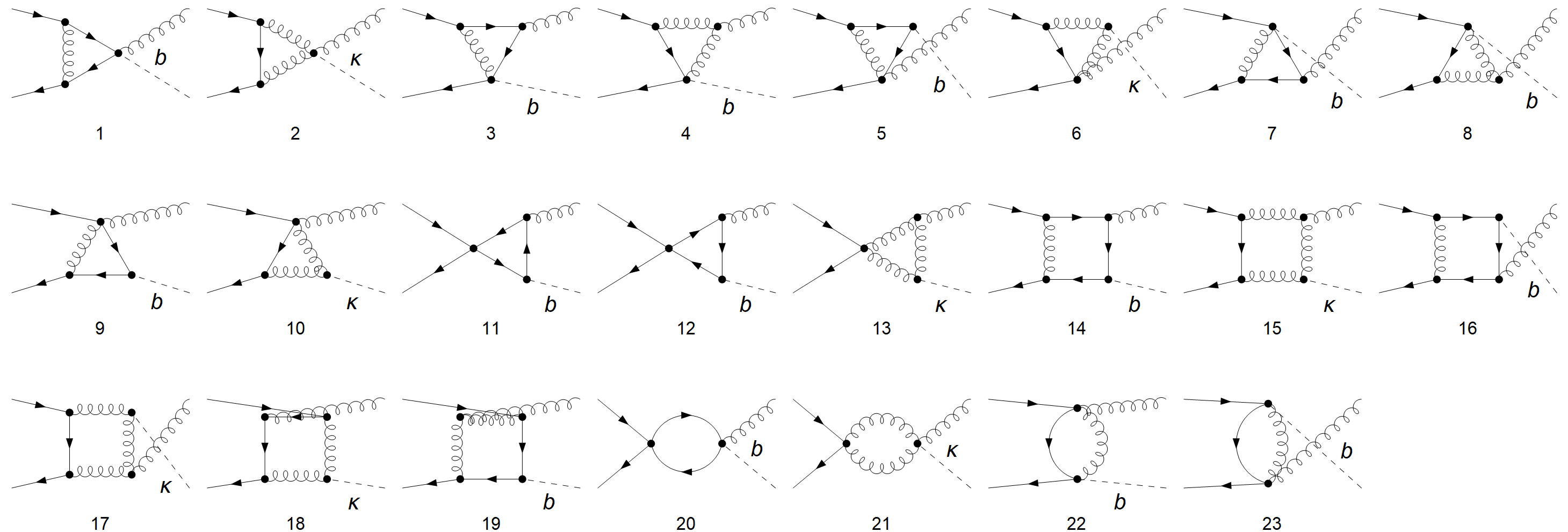}
    \caption{One-loop contributions to the scalar-scalar-gluon vertex at first order in the LV insertion.}
    \label{Fig15}
\end{figure}

The one-loop Feynman diagrams contributing to the scalar--scalar--gluon vertex with a single LV insertion are
displayed in fig.~\ref{Fig15}.

After summing all diagrams and isolating the UV pole, we obtain
\begin{equation}
\label{ssglv1}
\Gamma_{\mu}^{abc}(p_1,p_2)=
-\frac{Q_{1}g^{3}C_{A}\, (9 - 5\xi)}{64\, \pi^{2}\, \epsilon}
b_{\mu}\,
f^{abc}.
\end{equation}

\subsection{Scalar-Scalar-Gluon Counterterm}

\begin{figure}[h!]
    \centering
   \includegraphics[width=4cm, height=2cm]{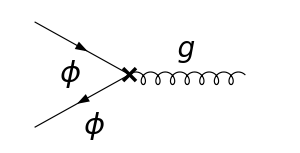}
    \caption{Counterterm for the Lorentz-invariant scalar-scalar-gluon vertex.}
    \label{SSG}
\end{figure}

\begin{figure}[h!]
    \centering
   \includegraphics[width=4cm, height=2cm]{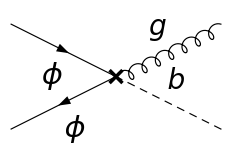}
    \caption{Counterterm for the scalar-scalar-gluon vertex at first order in the LV insertion.}
    \label{SSGLV}
\end{figure}

The renormalization of the scalar-scalar-gluon vertex requires
the introduction of counterterms to absorb the UV divergences. The counterterm vertices,
with and without the Lorentz violation, are shown in in figs.~\ref{SSG} and~\ref{SSGLV}.

From the divergent parts of the amplitudes in eqs.~\eqref{ssg1} and \eqref{ssglv1}, two
identical counter-terms may identified:
\begin{equation}
\label{eq:delta1}
\delta_{1}=\delta_{1_{Q_{1}}}=\frac{g^{2}C_{A}\,(9-5\xi)}{64\,\pi^{2}\,\epsilon}.
\end{equation}
Here, $\delta_{1}$ corresponds to the renormalization of the standard scalar-scalar-gluon vertex,
and $\delta_{1_{Q_{1}}}$ accounts for the LV modification. The equality of the two counterterms in
another consequence of gauge invariance.

\section{Calculations of the Two-Scalar-Two-gluon Function}
\label{sec-2phi-2g-vertices}

\subsection{Lorentz-Invariant Contributions}

\begin{figure}[h!]
    \centering
   \includegraphics[width=10cm, height=6cm]{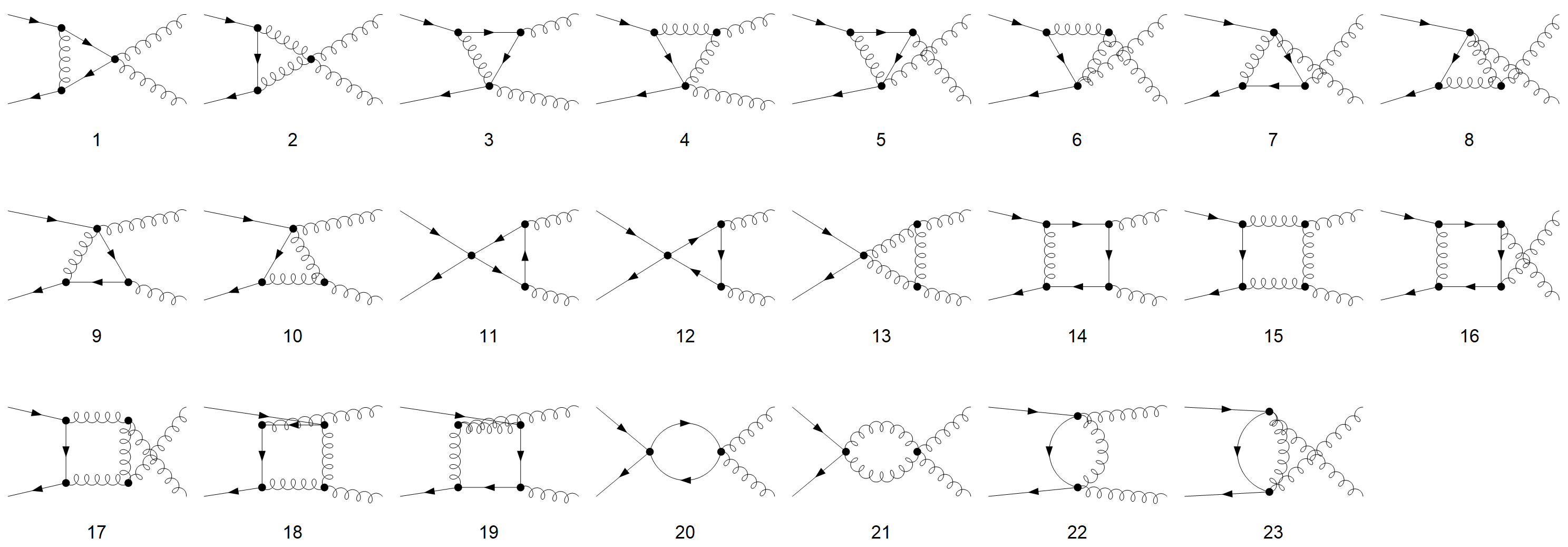}
    \caption{One-loop contributions to the Lorentz-invariant scalar-scalar-gluon-gluon vertex.}
    \label{Fig20}
\end{figure}

We now turn to the one-loop contribution to the scalar-scalar-gluon-gluon vertex. The corresponding Feynman
diagrams are all displayed in fig.~\ref{Fig20}. Retaining only the UV-divergent parts
(which may be found by neglecting all the external momenta, since the diagrams have only
a logarithmic superficial degree of divergence), we can write the
contribution associated with fig.~\ref{Fig20} as
\begin{eqnarray}
\Gamma_{\mu\nu}^{abcd}(p)&=&-\frac{3\,g^{4}C_{A}\, (\xi-1)}{16\pi^{2}\,\epsilon}\,g_{\mu\nu}
\left[
\mathrm{Tr}\!\left(T^{a}T^{b}T^{c}T^{d}\right)
+\mathrm{Tr}\!\left(T^{a}T^{b}T^{d}T^{c}\right)\right.
\nonumber\\
&&
-\,2\,\mathrm{Tr}\!\left(T^{a}T^{c}T^{b}T^{d}\right)
+\mathrm{Tr}\!\left(T^{a}T^{c}T^{d}T^{b}\right)
\nonumber\\
&&
-\,2\left.\mathrm{Tr}\!\left(T^{a}T^{d}T^{b}T^{c}\right)
+\mathrm{Tr}\!\left(T^{a}T^{d}T^{c}T^{b}\right)
\right].
\end{eqnarray}

\subsection{Contribution at First Order in the LV Insertions}

\begin{figure}[h!]
    \centering
   \includegraphics[width=14cm, height=10cm]{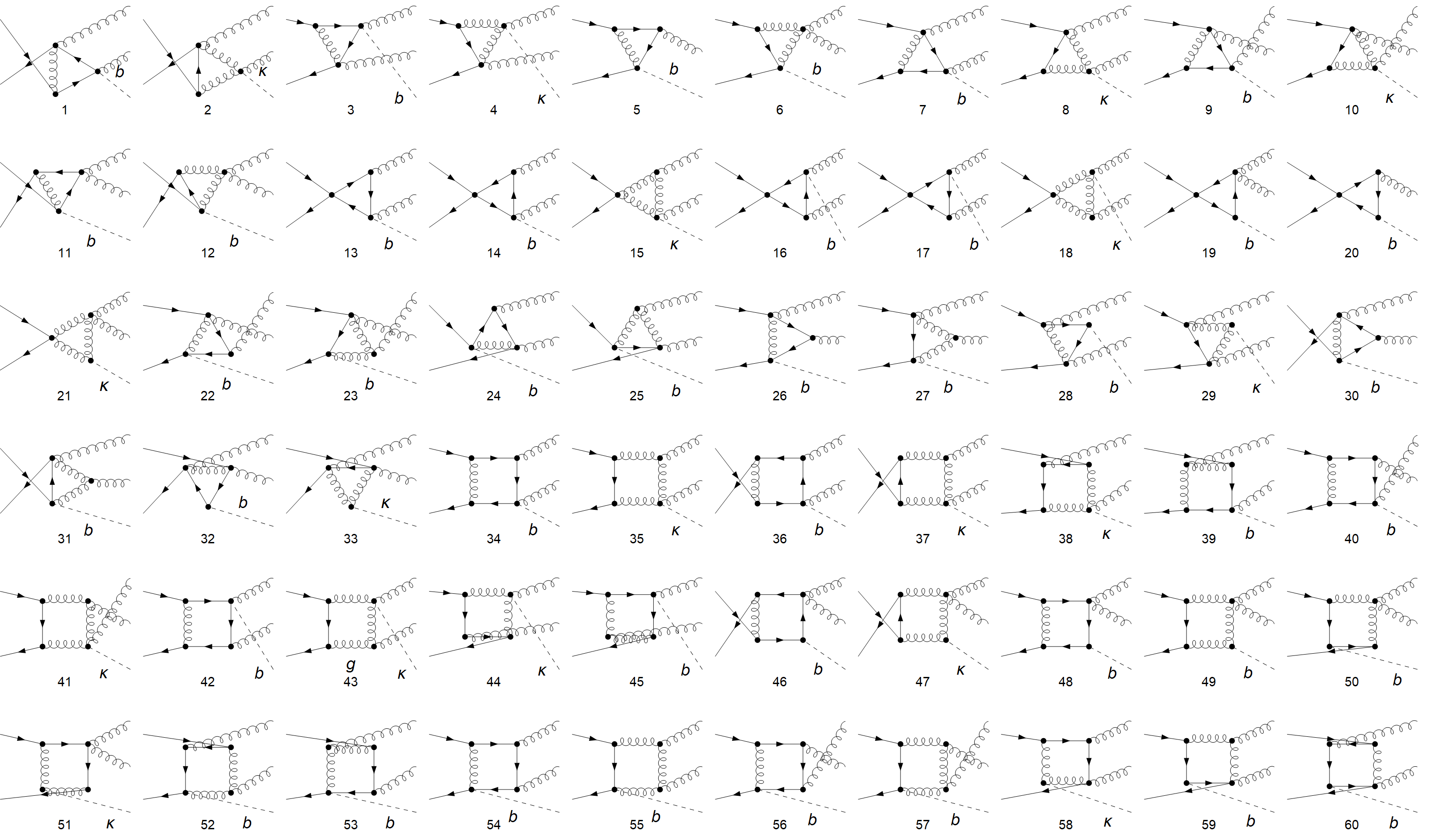}
    \caption{First set of one-loop contributions to the scalar-scalar-gluon-gluon vertex at first order in the
    LV insertion.}
    \label{Fig21}
\end{figure}

\begin{figure}[h!]
    \centering
   \includegraphics[width=14cm, height=10cm]{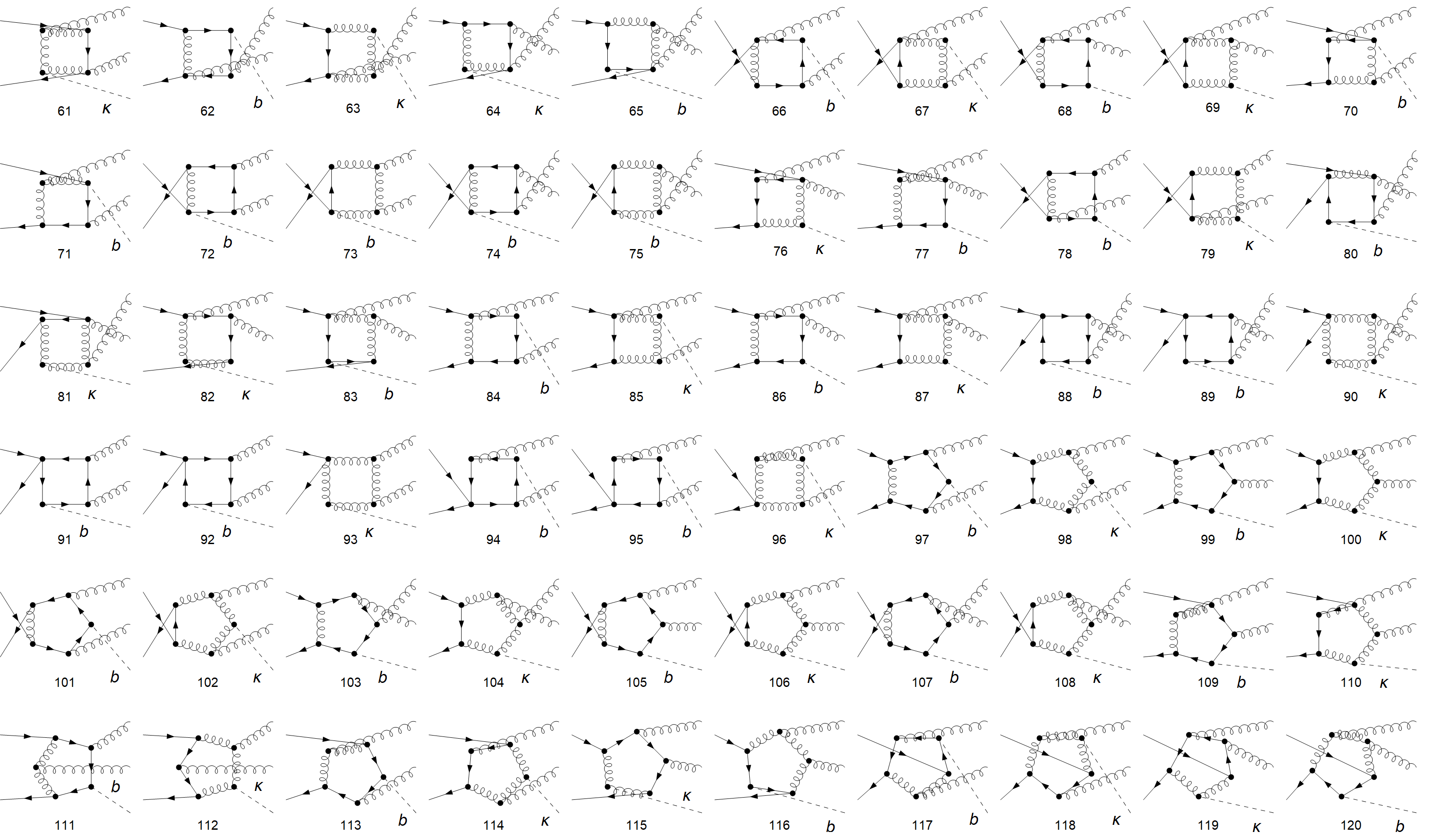}
    \caption{Second set of one-loop contributions to the scalar-scalar-gluon-gluon vertex at first order in the
    LV insertion.}
    \label{Fig22}
\end{figure}

The relevant contributions are represented by figs.~\ref{Fig21} and \ref{Fig22}. From an explicit
evaluation of these diagrams, we find that the sum of their UV-divergent parts vanishes identically;
this coincides with expectations based on power counting.
Therefore the corresponding one-loop contribution is UV finite at first order in the Lorentz violation.

\subsection{Counterterm}

The UV-divergent part of the one-loop correction to the $\phi^\dagger \phi A A$ vertex is
absorbed by the corresponding counterterm. From the result obtained above, we identify
\begin{equation}\label{eq:delta4}
\delta_{4}=\frac{3\,g^{2}C_{A}\,(1-\xi)}{32\pi^{2}\epsilon}.
\end{equation}
Therefore, the renormalization of the quartic scalar-gluon interaction is consistent
with multiplicative renormalizability for the theory.

\section{Calculation of the Scalar Self-Interaction Vertex}
\label{sec-4phi-vertex}

\begin{figure}[h!]
    \centering
   \includegraphics[width=10cm, height=6cm]{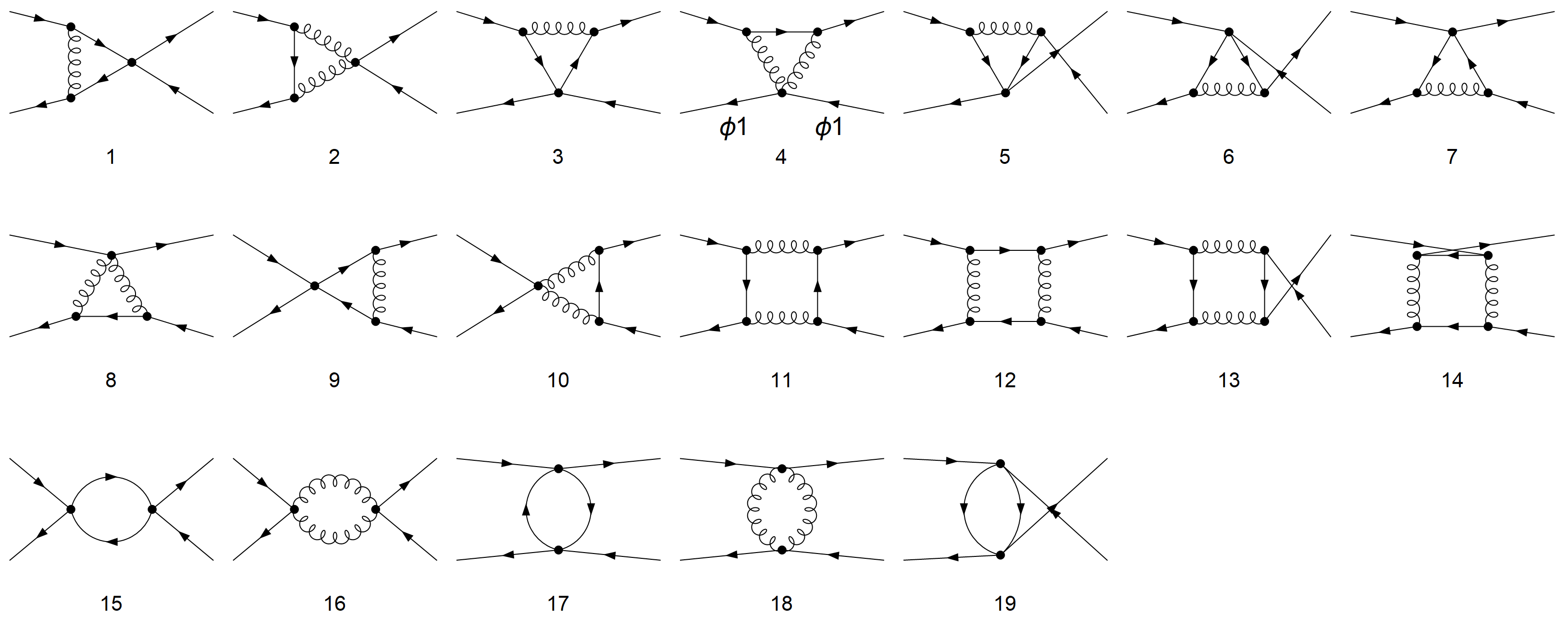}
    \caption{One-loop contributions to the four-scalar vertex.}
    \label{Fig28}
\end{figure}

The diagrams contributing to the scalar four-point function are shown in fig.~\ref{Fig28}.
After summing all diagrams and retaining only the divergent part, we find
\begin{eqnarray}
\mathcal{M}^{abcd} & = & 
-\frac{i}{64\,C_A^{2}\,\pi^{2}\epsilon}
\left[A_1\,\mathrm{tr}\!\left(T^{a}T^{b}T^{c}T^{d}\right)
+ A_2\,\mathrm{tr}\!\left(T^{a}T^{b}T^{d}T^{c}\right)
+ A_3\,\mathrm{tr}\!\left(T^{a}T^{c}T^{b}T^{d}\right)\right.
\nonumber\\
& & +\, A_4\,\mathrm{tr}\!\left(T^{a}T^{c}T^{d}T^{b}\right)
+ A_5\,\mathrm{tr}\!\left(T^{a}T^{d}T^{b}T^{c}\right)
+ A_6\,\mathrm{tr}\!\left(T^{a}T^{d}T^{c}T^{b}\right)
\label{Eq:Mtraces} \\
& & + \left.B_1\,\mathrm{tr}(T^{a}T^{d})
      \mathrm{tr}(T^{b}T^{c})
+ B_2\,\mathrm{tr}(T^{a}T^{c})
      \mathrm{tr}(T^{b}T^{d})
+ B_3\,\mathrm{tr}(T^{a}T^{b})
      \mathrm{tr}(T^{c}T^{d})\right], \nonumber
\end{eqnarray}
where the coefficients $A_{j}$ and $B_{j}$ are given by the expressions,
\begin{eqnarray}
\label{const1}
A_{1} & = & \left\{(3\lambda_2-5\lambda_3-2\lambda_4)(\lambda_2+\lambda_3+2\lambda_4)C_{A}^{2}
-12\lambda_1 (\lambda_2+\lambda_3+2\lambda_4)C_{A}
\right. \\
& & +\left. 8 \left(3\lambda_2\lambda_3 + 9\lambda_3^2 - 2\lambda_2\lambda_4 + 24\lambda_3\lambda_4 + 10\lambda_4^2\right)\right.\nonumber\\
&&\left.- 4 g^{2} \left[3g^{2}-2(\lambda_2+\lambda_3+2\lambda_4)\xi\right]C_{A}^{2}\right\}C_{A} ;
\nonumber\\
\label{const2}
A_{2} & = & -2\left\{
\left[5\lambda_2^2 - 6\lambda_2\lambda_3 + 5\lambda_3^2 + 4(\lambda_2+\lambda_3)\lambda_4 + 4\lambda_4^2\right]C_{A}^{2}
+12 \lambda_1(-\lambda_2+\lambda_3)C_{A}
\right. \\
& & + \left. 8\left[(3\lambda_2-7\lambda_3)\lambda_3 - 2(\lambda_2+5\lambda_3)\lambda_4 - 8\lambda_4^2\right]
+ 4 g^{2} \left[3g^{2}+2(\lambda_2-\lambda_3)\xi\right]C_{A}^{2}\right\}C_{A} ;
\nonumber
\end{eqnarray}
\begin{eqnarray}
\label{const3}
A_{3} & = & \left\{(3\lambda_2-5\lambda_3-2\lambda_4)(\lambda_2+\lambda_3+2\lambda_4)C_{A}^{2}
-12\lambda_1(\lambda_2+\lambda_3+2\lambda_4)C_{A}
\right. \\
& & +\left. 8\left(3\lambda_2\lambda_3+9\lambda_3^2-2\lambda_2\lambda_4+24\lambda_3\lambda_4+10\lambda_4^2\right)
-4g^{2}\left[3g^{2}-2(\lambda_2+\lambda_3+2\lambda_4)\xi\right]C_{A}^{2}\right\}\nonumber\\& &\times C_{A} ;
\nonumber
\end{eqnarray}
\begin{eqnarray}
\label{const4}
A_{4} & = &
-2\left\{\left[5\lambda_2^2 - 6\lambda_2\lambda_3 + 5\lambda_3^2 + 4(\lambda_2+\lambda_3)\lambda_4 + 4\lambda_4^2
\right]C_{A}^{2}+12\lambda_1(-\lambda_2+\lambda_3)C_{A}
\right. \\
& & +\left. 8\left[(3\lambda_2-7\lambda_3)\lambda_3 - 2(\lambda_2+5\lambda_3)\lambda_4 - 8\lambda_4^2\right]
+ 4g^{2}\left[3g^{2}+2(\lambda_2-\lambda_3)\xi\right]C_{A}^{2}
\right\}C_{A} ;
\nonumber\\
\label{const5}
A_{5} & = &  A_{6} \,\, = \,\,
\left\{(3\lambda_2-5\lambda_3-2\lambda_4)(\lambda_2+\lambda_3+2\lambda_4)C_{A}^{2}
-12\lambda_1(\lambda_2+\lambda_3+2\lambda_4)C_{A}
\right. \\
& & +\left. 8\left(3\lambda_2\lambda_3+9\lambda_3^2-2\lambda_2\lambda_4+24\lambda_3\lambda_4+10\lambda_4^2\right)
-4 g^{2}\left[3g^{2}-2(\lambda_2+\lambda_3+2\lambda_4)\xi\right]C_{A}^{2}
\right\}\nonumber\\
&&\times C_{A} ;
\nonumber\\
B_{1} & = &
-8\left[(\lambda_2+\lambda_3)^{2}C_{A}^{2}-12\lambda_1\lambda_{4}C_{A}
+4\left(\lambda_3^2+10\lambda_3\lambda_4+2\lambda_4^2\right)\right.\nonumber\\
&&\left.+4g^{2}(3g^{2}+2\lambda_4\xi)C_{A}^{2}
\right]; \\
\label{const8}
B_{2} & = &  B_{3} \,\, = \,\,
-4\left\{
\lambda_{1}^{2}C_{A}^{4}
+2\lambda_1(-\lambda_2+\lambda_3+2\lambda_4)C_{A}^{3}
+\left(3\lambda_1^2+2\lambda_2^2-2\lambda_3^2+4\lambda_4^2\right)C_{A}^{2}
\right. \\
& & - \left. 4\lambda_1(5\lambda_3+4\lambda_4)C_{A}
+44\lambda_3^2
+64\lambda_3\lambda_4
+32\lambda_4^2
+4g^{2}\left[6g^{2} + (-\lambda_{1}C_{A}+2\lambda_3)\xi\right]C_{A}^{2}\right\}.
\nonumber
\end{eqnarray}

\begin{figure}[h!]
    \centering
   \includegraphics[width=4cm, height=2cm]{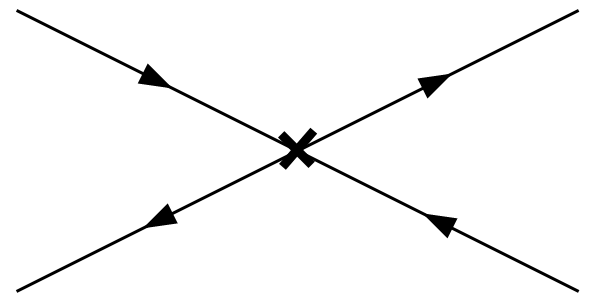}
    \caption{Counterterm for the four-scalar vertex.}
    \label{Fig42}
\end{figure}

In order to calculate the $\beta$-functions, we shall need the counterterms
corresponding to each $\lambda_{j}$, as well as the
field strength renormalization. After a lengthy calculation, we find
\begin{eqnarray}
\Gamma^{abcd}&  = &
i\Bigg[
\frac{8\,\delta\lambda_{4}}{C_{A}}\,
\mathrm{Tr}\!\left(T^{a}T^{d}\right)\,
\mathrm{Tr}\!\left(T^{b}T^{c}\right)
-2\,\delta\lambda_{1}\,
\mathrm{Tr}\!\left(T^{a}T^{c}\right)\,
\mathrm{Tr}\!\left(T^{b}T^{d}\right)
\nonumber\\
& &
+\frac{4\,\delta\lambda_{3}}{C_{A}}\,
\mathrm{Tr}\!\left(T^{a}T^{c}\right)\,
\mathrm{Tr}\!\left(T^{b}T^{d}\right)
-2\,\delta\lambda_{1}\,
\mathrm{Tr}\!\left(T^{a}T^{b}\right)\,
\mathrm{Tr}\!\left(T^{c}T^{d}\right)
\nonumber\\
& &
+\frac{4\,\delta\lambda_{3}}{C_{A}}\,
\mathrm{Tr}\!\left(T^{a}T^{b}\right)\,
\mathrm{Tr}\!\left(T^{c}T^{d}\right)
-\left(\delta\lambda_{2}+\delta\lambda_{3}+2\delta\lambda_{4}\right)
\mathrm{Tr}\!\left(T^{a}T^{b}T^{c}T^{d}\right) 
\nonumber\\
& &
+2\left(\delta\lambda_{2}-\delta\lambda_{3}\right)
\mathrm{Tr}\!\left(T^{a}T^{b}T^{d}T^{c}\right)
-\left(\delta\lambda_{2}+\delta\lambda_{3}+2\delta\lambda_{4}\right)
\mathrm{Tr}\!\left(T^{a}T^{c}T^{b}T^{d}\right)
\nonumber\\
& &
+2\left(\delta\lambda_{2}-\delta\lambda_{3}\right)
\mathrm{Tr}\!\left(T^{a}T^{c}T^{d}T^{b}\right)
-\left(\delta\lambda_{2}+\delta\lambda_{3}+2\delta\lambda_{4}\right)
\mathrm{Tr}\!\left(T^{a}T^{d}T^{b}T^{c}\right)
\nonumber\\
& &
-\left(\delta\lambda_{2}+\delta\lambda_{3}+2\delta\lambda_{4}\right)
\mathrm{Tr}\!\left(T^{a}T^{d}T^{c}T^{b}\right)
\Bigg],
\label{fsvertex}
\end{eqnarray}
where the expressions for the individual $\delta\lambda_{1}$, $\delta\lambda_{2}$,
$\delta\lambda_{3}$, and $\delta_{\lambda_{4}}$ are
\begin{eqnarray}
\label{lambda1}
\delta\lambda_1 & = &
\frac{1}{
32 C_A^2 \pi^2 \epsilon} 
\left\{
\lambda_{1}^{2}C_{A}^{4}
+2\lambda_1(-\lambda_2+\lambda_3+2\lambda_4)C_{A}^{3}
+\right.\nonumber\\
& & +\left.
\left[3\lambda_1^2+4(\lambda_2^2+\lambda_3^2+2\lambda_3\lambda_4+2\lambda_4^2)\right]C_{A}^{2}
\right.
\nonumber\\
& & -\left. 8\lambda_1(\lambda_3+2\lambda_4)C_{A}
-16(\lambda_3^2+2\lambda_3\lambda_4+2\lambda_4^2)
+4g^{2}(12g^{2}-\lambda_{1}C_{A}\xi)C_{A}^{2}
\right\};
\\
\label{lambda2}
\delta\lambda_2 & = &
\frac{1}{64 C_A \pi^2 \epsilon}
\left[
(-3\lambda_2^2+6\lambda_2\lambda_3+\lambda_3^2-4\lambda_2\lambda_4+4\lambda_3\lambda_4)C_{A}^{2}
+12 \lambda_1 \lambda_2 C_{A}
\right. 
\nonumber\\
& & -\left. 4\big(6\lambda_2\lambda_3+\lambda_3^2-4\lambda_2\lambda_4+4\lambda_3\lambda_4\big)
+4 g^{2}(3g^{2}-2\lambda_2)C_{A}^{2}
\right];\\
\label{lambda3}
\delta\lambda_3 & = &
\frac{1}{
32 C_A \pi^2 \epsilon}
\left[(\lambda_2^2+3\lambda_3^2+4\lambda_3\lambda_4+2\lambda_4^2)C_{A}^{2}
+6 \lambda_1 \lambda_3 C_{A}
\right.
\nonumber \\
& & 
-\left. 2(15\lambda_3^2+24\lambda_3\lambda_4+16\lambda_4^2)
+4 g^{2}(3g^{2}-\lambda_3\xi)C_{A}^{2}
\right];
\\
\label{lambda4}
\delta\lambda_4 & = &
-\frac{1}{64 C_A \pi^2 \epsilon}
\left[
(\lambda_2+\lambda_3)^{2}C_{A}^{2}
-12 \lambda_1\lambda_4 C_{A}
+4(\lambda_3^2+10\lambda_3\lambda_4+2\lambda_4^2)
\right.
\nonumber\\
& & +\left.
4g^{2}(3g^{2}+2\lambda_4\xi)C_{A}^{2}
\right].
\end{eqnarray}

A noteworthy (and potentially confusing) feature of the quartic scalar sector is the appearance of the adjoint Casimir
constant $C_A$ in the denominators of the $\delta\lambda_{j}$ counterterms.
This does not signal some kind of dynamical phenomenon that becomes singular in the Abelian limit $C_{A}=0$.
Rather, it is a consequence of the color decomposition used to extract the renormalization of the quartic couplings.
The couplings $\lambda_{j}$ were introduced in eq.~\eqref{LQCD-matrix} in a basis of gauge-invariant quartic operators
built from $\delta^{ab}$, $f^{abe}f^{cde}$, and $d^{abe}d^{cde}$ structures, whereas the one-loop four-scalar amplitude
is naturally organized in the trace basis, according to eq.~\eqref{Eq:Mtraces}. Matching the divergent amplitudes
to the counterterm vertices of eq.~\eqref{fsvertex} therefore requires inverting the mixing matrix between these two
bases of the color space. Since the normalization of the $\lambda_{j}$ basis depends on $C_{A}$, the resulting
expressions for the individual counterterms necessarily involve factors of $1/C_{A}$. Therefore, the denominators
in eqs.~\eqref{lambda1}--\eqref{lambda4} should be understood as group-theoretical projection
factors associated with the non-Abelian color algebra, rather than as signs of any pathological behavior.

Note that although we have written the counterterms in terms of the abstract quantity $C_{A}$, they have
only been verified for a $SU(N)$ gauge group. Moreover, there is a further subtlety to these results. The coupling
$\lambda_{2}$ does not admit an Abelian limit, as it multiplies a term in Lagrange density that involves
$f^{abc}$, which vanishes in the Abelian limit. This behavior arises from the fact that this couplings is
constructed from commutators of the fields. The other three couplings are also not all independent in the Abelian case.
The are built out of symmetric objects like anticommutators of the fields, but the definition of $d^{abc}$ is
ambiguous in a $U(1)^{N}$ theory. As a result, there are only two independent scalar couplings in the case of
an Abelian gauge multiplet. Under renormalization, all the structures in the non-Abelian theory become intertwined,
which accounts for the appearance of the parameter $C_{A}$ in eqs.~\eqref{Eq:Mtraces} and
\eqref{lambda1}--\eqref{lambda4}.

At this one-loop order and to the first order in the LV insertion, the four-scalar vertex receives no UV divergent
contribution. The sum of the divergent parts of all relevant diagrams vanishes identically, so that the
corresponding amplitude is UV finite in this sector, as should be expected from power counting.

\section{Calculations of the Ghost Field Self-Energy}
\label{sec-SE-ghost}

\subsection{Zeroth Order in the CFJ Parameter}

\begin{figure}[h!]
    \centering
   \includegraphics[width=4cm, height=4cm]{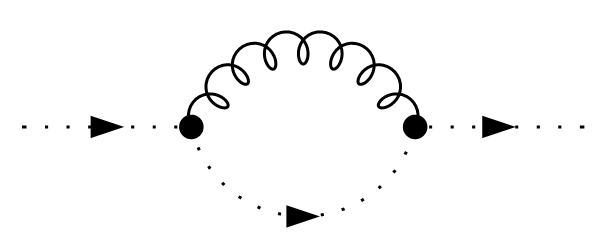}
    \caption{One-loop contribution to the ghost self-energy.}
    \label{Fig34}
\end{figure}
\noindent
The standard contribution to the ghost field propagator at zeroth order in the Lorentz violation is depicted in
fig.~\ref{Fig34}. After the calculation, we have the following result for the divergent part:
\begin{equation}
\Pi^{ab}(p^{2})=
-\frac{i\, g^{2}C_{A}\,(3-\xi)}
{64\, \pi^2\, \epsilon}\, p^2\,\delta^{ab}.
\end{equation}

\subsection{Vanishing First-Order Result}
\label{sec-SE-ghost-LV}

\begin{figure}[h!]
    \centering
   \includegraphics[width=4cm, height=4cm]{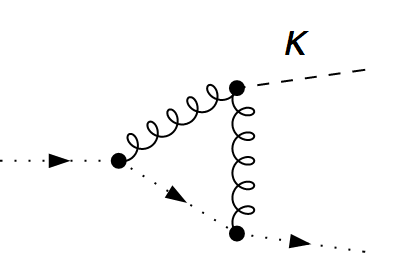}
    \caption{One-loop contribution to the ghost self-energy at first order in the LV insertion.}
    \label{Fig35}
\end{figure}

Considering the function with a Lorentz-violating CFJ insertion, we have the diagram shown in fig.~\ref{Fig35}.
Explicit calculations show that this three-point function is zero. The diagram is
finite by power counting, and the finite part also vanishes. This is actually not surprising, as there is no C-odd
two-ghost term that can be written down.
There is thus no contribution to the ghost effective action in the first order in the LV parameter.

\subsection{Counterterm for the Ghost Self-Energy}

\begin{figure}[h!]
    \centering
    \includegraphics[width=2cm,height=2cm]{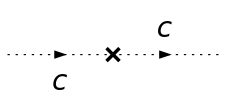}
    \caption{Counterterm for the ghost two-point function.}
    \label{Fig36}
\end{figure}

The one-loop counterterm associated with the ghost two-point function is represented diagrammatically in fig.~\ref{Fig36}.
By comparing the divergent part of the one-loop ghost self-energy with the corresponding tree-level structure,
we identify the usual form
\begin{equation}\label{eq:deltac}
\delta_{c}=\frac{g^{2}C_{A}(3-\xi)}{64\pi^{2}\epsilon}.
\end{equation}

\section{Calculations of the Ghost-Ghost-Gluon Vertex}
\label{sec-2c-g-vertex}

\subsection{Zeroth Order in the CFJ Term}

\begin{figure}[h!]
    \centering
   \includegraphics[width=8cm, height=4cm]{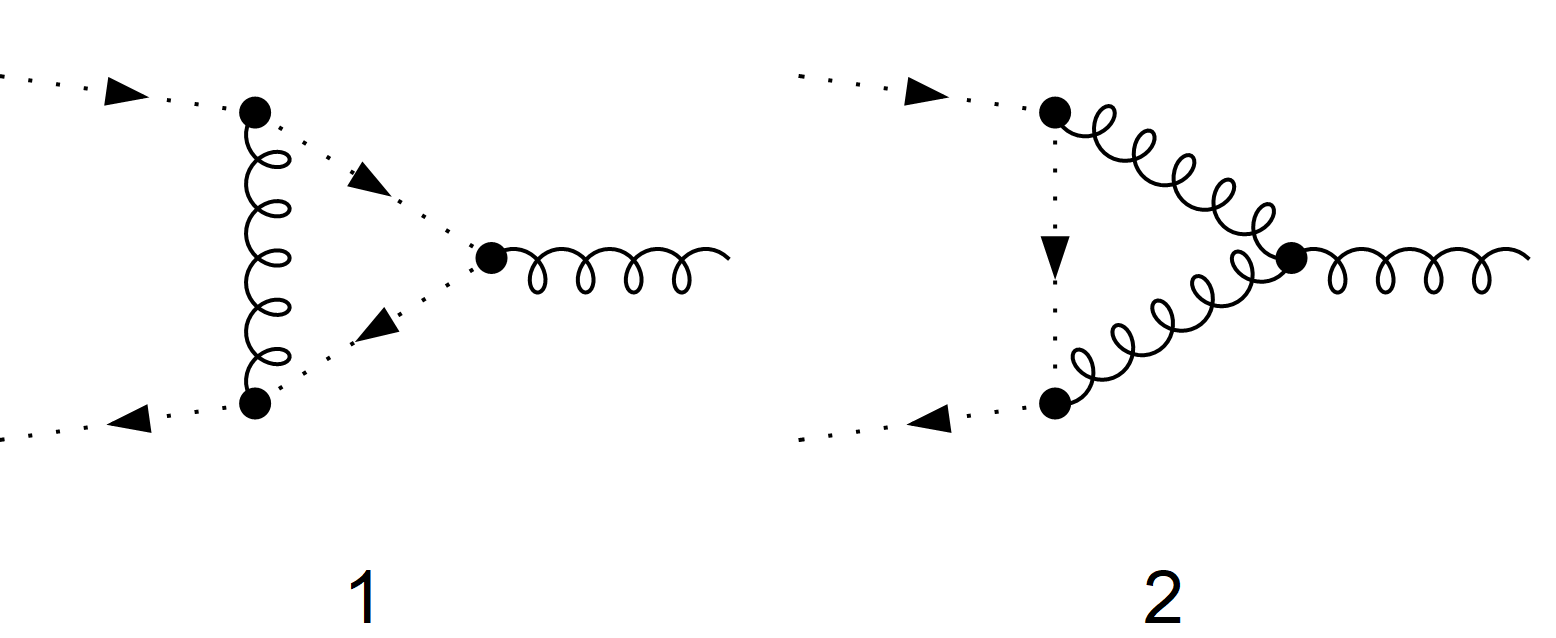}
    \caption{One-loop contribution to the ghost-ghost-gluon vertex.}
    \label{Fig37}
\end{figure}

Finally, we compute the radiative corrections to the ghost-ghost-gluon interaction vertex.
At zeroth order in the gauge-sector parameter $\kappa_{\mu}$, we have the standard
pair of diagrams shown in fig.~\ref{Fig37}.
After some calculation, we have the result,
\begin{equation}
\Pi_{\mu}^{abc}(p_{2})=
\frac{g^{3}C_{A} \, \xi}{32 \, \pi^2 \, \epsilon}\,p_{2\mu}\,f^{a b c} .
\end{equation}

\subsection{Vertex at First Order in the LV parameter}

\begin{figure}[h!]
    \centering
   \includegraphics[width=10cm, height=4cm]{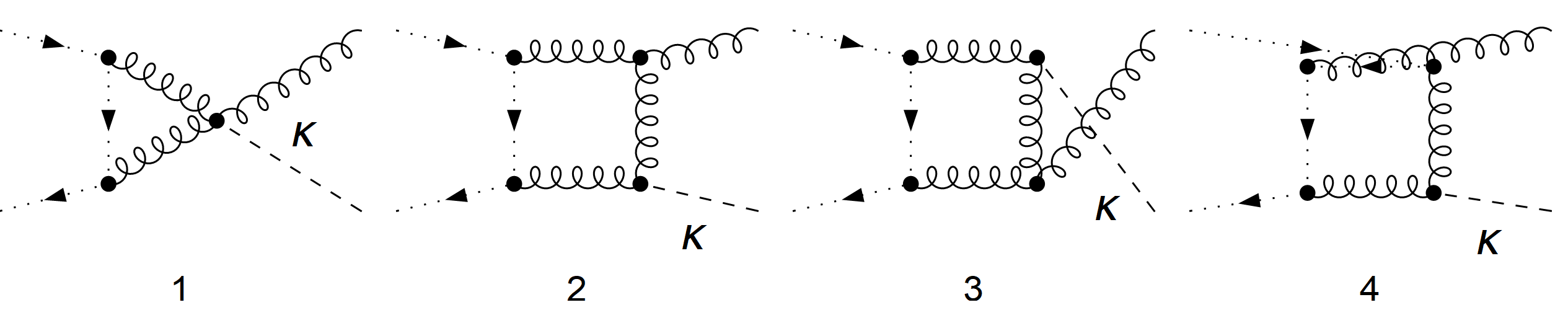}
    \caption{One-loop contributions to the ghost-ghost-gluon vertex at first order in the LV insertion.}
    \label{Fig38}
\end{figure}

At first order in LV parameter, we have the diagrams depicted in fig.~\ref{Fig38}.
After calculation, the sum of these contribution is zero, which implies that there are no contributions
to the ghost-ghost-gluon-coupling at first order in the LV CFJ parameter. This is unsurprising, for the same
symmetry reasons as in section~\ref{sec-SE-ghost-LV}.

\subsection{Couterterm for the Vertex}

\begin{figure}[h!]
    \centering
   \includegraphics[width=4cm, height=2cm]{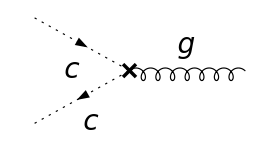}
    \caption{Counterterm for the ghost-ghost-gluon vertex.}
    \label{Fig39}
\end{figure}

The counterterm to ghost-ghost-gluon contribution can be seen in fig.~\ref{Fig39}.
and we identify this counterterm as the usual one,
\begin{equation}
\label{eq:delta1c}
\delta_{1c} \;=\;
-\frac{g^{2}C_{A} \, \xi}{
32 \, \pi^2 \, \epsilon},
\end{equation}
which plays a role in the determination of the renormalization group $\beta$-functions of the theory.


\begin{table*}[t]
\centering
\caption{Summary of the one-loop renormalization counterterm for the model.}
\label{tab:counterterms}
\begin{tabular}{ccc}
\hline\hline
Renormalization constant & Meaning & Formula \\
\hline
$\delta_{3}$ & gluon wave-function counterterm & \eqref{eq:delta3} \\
$\delta_{2_{CFJ}}$ & quadratic CFJ-sector counterterm & \eqref{2CFJ} \\
$\delta_{3g}$ & three-gluon counterterm & \eqref{eq:delta3g} \\
$\delta_{3_{CFJ}}$ & cubic CFJ-sector counterterm & \eqref{3CFJ} \\
$\delta_{4g}$ & four-gluon counterterm & \eqref{eq:delta4g} \\
$\delta_{2}$ & scalar wave-function counterterm & \eqref{eq:delta2} \\
$\delta_{Q_{1}}$ & scalar LV two-point counterterm & \eqref{eq:deltaQ1} \\
$\delta_1$ & scalar-scalar-gluon counterterm & \eqref{eq:delta1} \\
$\delta_{1_{Q_{1}}}$ & scalar-scalar-gluon LV counterterm & \eqref{eq:delta1} \\
$\delta_4$ & two-scalar-two-gluon counterterm & \eqref{eq:delta4} \\
$\delta_{\lambda_1}$ & quartic scalar counterterm & \eqref{lambda1} \\
$\delta_{\lambda_2}$ & quartic scalar counterterm & \eqref{lambda2} \\
$\delta_{\lambda_3}$ & quartic scalar counterterm & \eqref{lambda3} \\
$\delta_{\lambda_4}$ & quartic scalar counterterm & \eqref{lambda4} \\
$\delta_c$ & ghost wave-function counterterm & \eqref{eq:deltac} \\
$\delta_{1c}$ & ghost--ghost--gluon counterterm & \eqref{eq:delta1c} \\
\hline\hline
\end{tabular}
\end{table*}


\section{Renormalization Group $\beta$-Functions}
\label{sec-beta-functions}

The counterterms obtained in the previous sections---and summarized in table~\ref{tab:counterterms}---enable
us to complete the one-loop renormalization program by extracting the corresponding renormalization group
$\beta$-functions, which determine the scale dependence of the interactions processes as a function of the invariant
momentum exchange, well above the scale of the scalar mass.
We shall look explicitly at the the one-loop running of quantities in the gauge sector, the CPT-odd LV operators,
and the multiple scalar self-interactions. For the gauge coupling, we find
\begin{equation}
    \beta_{g}=-\frac{10\,g^{3}C_{A}}{96\pi^{2}}.
\end{equation}
The 10 in the numerator represents $(11-N_{s})$;
the contribution to this $\beta$-function from the scalar field is positive, while the contribution from the gauge
sector (including ghost contributions) is negative and 11 times larger in magnitude. The scalar
contribution is analogous to the better-known behavior of spinor matter in QCD.

In the LV sector, the $\beta$-functions are found to be
\begin{equation}
    \beta_{Q_{1}}=0
\end{equation}
and
\begin{equation}
    \beta_{Q_{2}}=-\frac{10\,Q_{2}g^{2}C_{A}}{48\pi^{2}}.
\end{equation}
at this order.
Accordingly, for the combination $Q_{2}g$ we obtain
\begin{equation}
    \beta_{Q_{2}g}=-\frac{10\,Q_{2}g^{3}C_{A}}{32\pi^{2}}.
\end{equation}

Finally, turning to the $\lambda$ sector, the corresponding $\beta$-functions are given by
\begin{eqnarray}
\beta_{\lambda_{1}} & = &
\frac{\lambda_{1}\left[\lambda_{1}(3+C_{A}^{2})+2(-\lambda_{2}+\lambda_{3}+2\lambda_{4})C_{A}\right]
+4(\lambda_2^2+\lambda_3^2+2\lambda_3\lambda_4+2\lambda_4^2)}
{16\pi^2}
\nonumber \\
& &+\frac{3g^{2}(4g^{2}-\lambda_{1}C_{A})}{4\pi^2} -\frac{\lambda_{1}(\lambda_3+2\lambda_4)}{2C_{A}\pi^2}
-\frac{\lambda_3^2+2\lambda_3\lambda_4+2\lambda_4^2}{C_A^2\pi^2}
\\
\beta_{\lambda_{2}} & = &
\frac{\lambda_{2}\left[(-3\lambda_{2}+6\lambda_{3}-4\lambda_{4})C_{A}+12\lambda_{1}\right]
+(\lambda_3^2  + 4 \lambda_3\lambda_4)C_{A}}{32 \pi^2}
\nonumber \\
& & +\frac{3g^{2}(g^{2}-2\lambda_{2})C_{A}}{8 \pi^2}
-\frac{2\lambda_{2}(3\lambda_{3}-4\lambda_{4})+\lambda_3^2 + 4\lambda_3\lambda_4}{8 C_{A} \pi^2}
\\
\beta_{\lambda_{3}} & = &
\frac{\lambda_{3}\left[(3\lambda_{3}+4\lambda_{4})C_{A}+6\lambda_{1}\right]+(\lambda_2^2+2\lambda_4^2)C_{A}}{16\pi^2}
+\frac{3g^{2}(g^{2}-\lambda_{3})C_{A}}{4\pi^2}
\nonumber\\
& & -\frac{3\lambda_{3}(5\lambda_{3}+8\lambda_{4})+16\lambda_{4}^{2}}
{8C_{A}\pi^2}
\\
\beta_{\lambda_{4}} & = &
\frac{12\lambda_{4}\lambda_{1}-(\lambda_2+\lambda_3)^{2}C_{A}}{32 \pi^2}
-\frac{3g^{2}(g^{2}+2\lambda_{4})C_{A}}{8\pi^2}
-\frac{2\lambda_{4}(\lambda_{4}+5\lambda_{3})+\lambda_{3}^{2}}{8 C_A \pi^2}.
\end{eqnarray}
Note that these are all gauge invariant, not depending on $\xi$.

The $\beta$-functions derived above complete the one-loop renormalization analysis of the model and confirm
the internal consistency of its perturbative structure. In particular, the LV couplings exhibit a 
renormalization pattern fully compatible with multiplicative renormalizability, since their running is
entirely governed by counterterms of the same form as those appearing in the original Lagrange density.
At the same time, the resulting renormalization group equations make transparent the interplay between the
gauge, scalar, and CPT-odd sectors, thereby providing the appropriate framework for future studies of the
UV flow and possible asymptotic regimes of the theory. We note that for certain relations between couplings,
some of $\beta$-functions vanish at this order, which in principle could signalize the existence of fixed points.

\section{Summary}
\label{sec-summary}

In this work, we carried out a systematic one-loop renormalization analysis of CPT-odd LV scalar QCD.
Treating the LV operators as perturbative insertions, we evaluated the UV-divergent contributions to the
relevant two-, three-, and four-point Green's functions in both the gauge and scalar sectors, including the
ghost contributions whenever required. Our results show that the divergent part of the gluon two-point function
at first order in the background vector reproduces the expected CFJ structure, while the scalar two-point
function generates the corresponding CPT-odd contribution proportional to $b\cdot p$. We further determined the
divergent corrections to the three-gluon, scalar-scalar-gluon, and quartic interaction vertices and extracted the
associated counterterms.

A central result of this analysis is that, as expected, all the one-loop divergences can be absorbed into counterterms
already present in the classical action. In this way, we have explicitly verified the multiplicative renormalizability
of the model at one-loop order (except for the scalar mass term, which would be additively renormalized).
In particular, the counterterms obtained for the Lorentz-invariant and LV sectors consistently reproduce the
tensor, color, and derivative structures dictated by the original Lagrangian. We also reconfirm the
cancellation of the UV-divergent part of the four-gluon vertex with a single LV insertion,
in agreement with the general expectations based on power counting.

The triviality of the $\beta$-function for the CPT-violating term in the scalar sector is not unexpected.
The bilinear part of this term---which has the form of an interaction between a charged scalar and a constant,
nondynamical vector potential (meaning a theory with a vanishing Abelian field strength)---has
previously been analyzed in the presence of a additional interactions~\cite{ref-altschul44}. A field
redefinition removes this term from the action; the redefintion is equivalent in the context of
quantum corrections to a simple translation of the integrated loop momentum, and the term generally has
no observable effects, even nonperturbatively, for a single scalar field or when multiple scalar fields
are coupled identically to the constant vector potential background. However, when multiple particle
species interact with multiple backgrounds, the differences between the couplings in different sectors may become
observable. (All of these observations are also true of the analogous type term for fermions, the
SME $a_{\mu}$~\cite{KosPic,Colladay:2007aj,ref-altschul44,ref-kost6}.)

A further remark concerns the qualitative behavior of the renormalization-group flow. Since the one-loop gauge beta function
is proportional to $-(11-N_{s})\,C_{A}$, the model is asymptotically free in the gauge sector for $N_{s}<11$, while for
$N_{s}>11$ the running is reversed and the gauge coupling becomes IR free. This mirrors the behavior in spinor QCD, where
if there are enough generations of quarks, the theory coupling constant grows at momentum exchanges above the largest quark
mass.  The borderline case $N_{s}=11$ in the scalar theory yields a vanishing one-loop $\beta$-function for the gauge
coupling $g$ (although, unlike in theories with extended supersymmetry) there is no particuarly reason to expect that the
$\beta$-function will continue to vanish at $\mathcal{O}(\hbar^{2})$ and higher orders.

The same overall factor controls the running of the CPT-odd parameter $Q_2$ and the combination $Q_{2}g$,
while $Q_1$ does not run at this order. Therefore, the UV or IR behavior of the LV sector is tightly correlated with
that of the pure non-Abelian gauge interaction. The scalar sector is qualitatively more involved. Since the
$\beta$-functions for $\lambda_1$, $\lambda_2$, $\lambda_3$, and $\lambda_4$ generate a coupled nonlinear system
of differential equations,
their UV behaviors cannot be characterized by a simple sign criterion analogous to that
applying to $\beta_g$. Instead,
the renormalization-group flow of the scalar self-interactions must be analyzed in the full multidimensional
parameter space. In particular, the forms of the $\beta$-functions leaves open the possibility of nontrivial
fixed points for specific relations among the couplings.

The present results provide the non-Abelian CPT-odd counterpart of earlier studies on LV scalar gauge theories
and establish a consistent perturbative framework for further investigations. Natural continuations of this work
would include detailed higher loop calculations,
study of the mass renormalization, characterization of the finite parts of the amplitudes,
analysis of possible implications for spontaneous symmetry breaking, explicit determination of the effective potential,
and studies of other nontrivial dynamical phenomena in LV non-Abelian scalar theories.
Besides this, it is worth mentioning that, since scalar QED may serve as a toy model within studies of
gravitational scattering~\cite{Bern:2021xze}, it is natural to expect that our results could also
be applied for studies of
LV effects in gravitational scattering. We plan to perform such studies in forthcoming papers.

\section*{Acknowledgements}

The work of A. Yu.\ P. has been partially supported by the CNPq project no. 303777/2023-0.
The work of A. C. L was partially supported by CNPq, grants no.\ 404310/2023-0 and no.\ 301256/2025-0.

\bibliographystyle{plain}

\end{document}